\documentclass[10pt,twocolumn]{IEEEtran}
\makeatletter\newif\if@restonecol

\usepackage{graphicx}
\usepackage{epstopdf}
\usepackage{eucal}
\usepackage{times}
\usepackage{latexsym}
\usepackage{amsmath}
\usepackage{amssymb}
\usepackage{amsthm}
\usepackage{epsfig}
\usepackage{cases}
\usepackage{amsfonts}
\usepackage{indentfirst}
\usepackage{amsopn}
\usepackage{bm}
\usepackage{algorithmic}
\usepackage[linesnumbered,ruled]{algorithm2e}
\usepackage{subfigure}
\usepackage{stfloats}
\usepackage{booktabs}
\usepackage{slashbox}
\usepackage{arydshln}
\usepackage{amssymb}
\usepackage{color}
\usepackage{pifont}
\usepackage{multirow}
\usepackage[justification=centering]{caption}
\usepackage[backend=bibtex, sorting=none, citestyle=numeric-comp, style=ieee]{biblatex}

\newenvironment{shrinkeq}[1]%
{ \bgroup
  \addtolength\abovedisplayshortskip{#1}
  \addtolength\abovedisplayskip{#1}
  \addtolength\belowdisplayshortskip{#1}
  \addtolength\belowdisplayskip{#1}}
{\egroup\ignorespacesafterend}

\usepackage{xcolor}
\DeclareBibliographyCategory{important}
\colorlet{impentry}{blue}

\AtEveryBibitem{%
  \ifcategory{important}%
    {\bfseries\color{impentry}}%
    {}%
  }

\addtocategory{important}{
}

\newtheorem{lemma}{Lemma}
\newtheorem{theorem}{Theorem}

\newtheorem{remark}{Remark}[section]

\makeatletter
\newcommand*\rel@kern[1]{\kern#1\dimexpr\macc@kerna}
\newcommand*\widebar[1]{%
  \begingroup
  \def\mathaccent##1##2{%
    \rel@kern{0.8}%
    \overline{\rel@kern{-0.8}\macc@nucleus\rel@kern{0.2}}%
    \rel@kern{-0.2}%
  }%
  \macc@depth\@ne
  \let\math@bgroup\@empty \let\math@egroup\macc@set@skewchar
  \mathsurround\z@ \frozen@everymath{\mathgroup\macc@group\relax}%
  \macc@set@skewchar\relax
  \let\mathaccentV\macc@nested@a
  \macc@nested@a\relax111{#1}%
  \endgroup
}

\begin{document}

\title{Intelligent Reflecting Surface Aided MISO Uplink Communication Network: Feasibility and Power Minimization for Perfect and Imperfect CSI}


\author{Yang Liu,\thanks{Y. Liu is with the School of Information and Communication Engineering, Dalian University of Technology, Dalian, China, email: yangliu\_613@dlut.edu.cn.}\ \ 
Jun Zhao, \emph{Member, IEEE},\thanks{J. Zhao is with the School of Computer Science and Engineering, Nanyang Technological University, Singapore, email: junzhao@ntu.edu.sg. He is supported by 1) Nanyang Technological University (NTU) Startup Grant, 2) Alibaba-NTU Singapore Joint Research Institute (JRI), 3) Singapore Ministry of Education Academic Research Fund Tier 1 RG128/18, Tier 1 RG115/19, Tier 1 RT07/19, Tier 1 RT01/19, Tier 1 RG24/20, and Tier 2 MOE2019-T2-1-176, 4) NTU-WASP Joint Project, 5) Singapore NRF National Satellite of Excellence, Design Science and Technology for Secure Critical Infrastructure NSoE DeST-SCI2019-0012, 6) AI Singapore (AISG) 100 Experiments (100E) programme, and 7) NTU Project for Large Vertical Take-Off \& Landing (VTOL) Research Platform. J. Zhao is the \emph{Corresponding author}.}\ \  
Ming Li, \emph{Senior Member, IEEE},\ and\thanks{M. Li is with the School of Information and Communication Engineering, Dalian University of Technology, Dalian 116024, China, and also with the National Mobile Communications Research Laboratory, Southeast University, Nanjing 210096, email: mli@dlut.edu.cn. His work is supported partly by the National Natural Science Foundation of China (Grant No. 61971088) and partly by the Open Research Fund of National Mobile Communications Research Laboratory, Southeast University (No. 2021D08).}\ \ 
Qingqing Wu, \emph{Member, IEEE}\thanks{Q. Wu is with the State Key Lab. of Internet of Things for Smart City, University of Macau, Macau, China, and also with the National Mobile Communications Research Laboratory, Southeast University, Nanjing 210096, email: qingqingwu@um.edu.mo. His work is supported by the Open Research Fund of National Mobile Communications Research Laboratory, Southeast University (No. 2021D15).}}


\maketitle

\begin{abstract}
In this paper, we consider the weighted sum-power minimization under quality-of-service (QoS) constraints in the multi-user multi-input-single-output (MISO) uplink wireless network assisted by intelligent reflecting surface (IRS). We perform a comprehensive investigation on various aspects of this problem. First, when users have sufficient transmit powers, we present a new sufficient condition guaranteeing arbitrary information rate constraints. This result strengthens the feasibility condition in existing literature. Then, we design novel penalty dual decomposition (PDD) based and nonlinear equality constrained alternative direction method of multipliers (neADMM) based solutions to tackle the IRS-dependent-QoS-constraints, which effectively solve the feasibility check and power minimization problems. Besides, we further extend our proposals to the cases where channel status information (CSI) is imperfect and develop an online stochastic algorithm that satisfy QoS constraints stochastically without requiring prior knowledge of CSI errors. Extensive numerical results are presented to verify the effectiveness of our proposed algorithms. 
\end{abstract}


\section{Introduction}
\label{sec:introduction}

\subsection{Background}
\label{subsec:survey}

The intelligent reflecting surface (IRS) has been envisioned as a key technology for the next generation wireless network \cite{QQWu_Mag2020}. It is constructed based on the \emph{metasurface} and possesses many appealing advantages---energy and cost efficiency, easiness for deployment and so on. IRS has been included in the white paper \cite{6G_WhitePaper} for 6G wireless network.

Recently a great number of research work has been made to explore IRS' potentials in enhancing the wireless communication networks \cite{QQWu_TWC2019, QQWu_discrete_TCom2020, CWHuang_TWC2019, WeiXu_IRS_WCLett, CunhuaPan_Multicell_IRS, IRS_ADMM_YCLiang2019, IRS_PWT_Ng, MengHua_IRS_Comp, QQWu_IRS_SWIPT2019, QQWu_CommLett2020, YifeiYang_OFDM_IRS, BZheng_IRS_Noma, IRS_ADMM_QiZhang2019, AI_IRS_2020, WeiXu_IRS_CommLett, Guan_CommLett2020, CunhuaPan_MEC_IRS, QUA_asymptotic_mmSINR, MMZhao_TTS, MMZhao_ImprctCSI, GuiZhou_Rbst_IRS_TSP, GuiZhou_Rbst_IRS_Lett, IRS_ChnEst_LDai, IRS_ChnEst_ZhaoruiWang, IRS_ChnEst_JieChen, IRS_practical_model}. In \cite{QQWu_TWC2019} the authors consider the power minimization problem in an IRS-assisted MISO multi-user downlink system via jointly optimizing active beamformers and IRS elements. Besides demonstrating the improvement in power saving, \cite{QQWu_TWC2019} also uncovers the squared power scaling law $N^2$ ($N$ is the number of reflecting units of IRS). More importantly, it is proved in \cite{QQWu_discrete_TCom2020} that such a fundamental power scaling law also holds for practical IRS with low-resolution such as $1$-bit phase shifters, which shows the great potential of using low-cost IRS in practice. \cite{CWHuang_TWC2019, WeiXu_IRS_WCLett} demonstrate that the energy efficiency (EE) can be significantly boosted via utilizing IRS in a multi-user MISO system. \cite{CunhuaPan_Multicell_IRS, IRS_ADMM_YCLiang2019} have shown that by collaboratively designing precoders at multiple base stations (BSs) together with the IRS elements in a downlink multicell scenario, the weighted sum rate of all cell-edge users can be greatly promoted. \cite{IRS_PWT_Ng} optimizes the sum-rate of a downlink MISO system with self-sustainable IRS, where energy harvesting and power consumption (due to reflecting elements' phase shifter) at the IRS is taken into account. \cite{MengHua_IRS_Comp} considers cooperative beamforming in joint processing coordinated multipoint downlink transmission and shows that with the help of IRS, the minimal rate of cell-edge users can be significantly improved. \cite{QQWu_IRS_SWIPT2019, QQWu_CommLett2020} perform a joint IRS and transmit beamforming optimization in a downlink simultaneous wireless information and power transfer (SWIPT) setting towards optimizing total power consumption or energy transferring while guaranteeing the quality-of-service (QoS) requirements for all information and energy users. Besides, the IRS has also been found of great potentials to benefit wireless networks under various settings. For instance, the recent works \cite{YifeiYang_OFDM_IRS, IRS_ADMM_QiZhang2019,BZheng_IRS_Noma, AI_IRS_2020, WeiXu_IRS_CommLett, Guan_CommLett2020, CunhuaPan_MEC_IRS} have illustrated the significance of IRS in improving system's performance in orthogonal frequency division multiplexing (OFDM), non-orthogonal multiple access (NOMA), physical layer secure transmission, cognitive radio (CR) and mobile edge computing (MEC), respectively. All the above works assume perfect channel status information (CSI). \cite{QUA_asymptotic_mmSINR} uses spatial correlation statistics to analyse the asymptotic performance of the large-scale MISO downlink system. \cite{MMZhao_TTS, MMZhao_ImprctCSI} consider using discrete-resolution IRS to boost weighted sum-rate for downlink MISO users when CSI is imperfect. \cite{GuiZhou_Rbst_IRS_TSP, GuiZhou_Rbst_IRS_Lett} perform robust beamforming design to reduce power consumption when CSI estimation has errors. Besides, a bunch of works \cite{IRS_ChnEst_LDai, IRS_ChnEst_ZhaoruiWang, IRS_ChnEst_JieChen} have researched jointly designing pilot sequences and IRS reflection patterns to perform low-overhead channel estimation. \cite{IRS_practical_model} considers a practical reflection model of IRS reflecting elements, where the amplitude response of the reflecting element is non-linearly coupled with the phase response.

\begin{figure}[!t]
\centering
\includegraphics[scale=0.27]{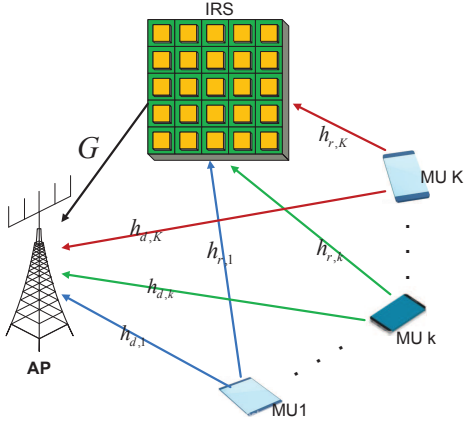}
\centering
\caption{An IRS-aided MISO uplink communication system.} \label{fig:fig-system}
\end{figure}

\subsection{Contributions}
\label{subsec:contribution}

This paper considers weighted sum-power minimization by jointly configuring IRS and allocating power in a MISO uplink multi-user system. This problem is rather challenging due to the presence of IRS-dependent-QoS-constraints, which is not considered in most of the literature \cite{CWHuang_TWC2019, WeiXu_IRS_WCLett, CunhuaPan_Multicell_IRS, IRS_ADMM_YCLiang2019, IRS_PWT_Ng, YifeiYang_OFDM_IRS, BZheng_IRS_Noma, AI_IRS_2020, CunhuaPan_MEC_IRS, QUA_asymptotic_mmSINR, MMZhao_TTS, MMZhao_ImprctCSI, WeiXu_IRS_CommLett, IRS_ChnEst_LDai, IRS_ChnEst_ZhaoruiWang, IRS_ChnEst_JieChen, IRS_practical_model}. To attack this problem, we propose novel methods that cherish advantageous convergence or efficiency properties compared to the very few methods dealing this problem in the existing literature \cite{QQWu_TWC2019,QQWu_discrete_TCom2020, Guan_CommLett2020, QQWu_CommLett2020, MengHua_IRS_Comp, GuiZhou_Rbst_IRS_TSP, GuiZhou_Rbst_IRS_Lett, IRS_ADMM_QiZhang2019}. Specific contributions are detailed as follows:

\begin{itemize}

\item[i)] First, we explore the feasibility of the QoS-constrained problem when transmitters have sufficient (unbounded) powers. This problem, i.e. the feasibility region characterization, is a classical one and has been extensively researched \cite{multiuser_MSchubert2005} and perfectly solved \cite{complete_QoS_RHunger} when IRS is absent. For IRS-aided system, however, the feasibility is an open issue. We partially answers this question via deriving a new sufficient condition, which strengthens the existing result in the literature, e.g. \cite{QQWu_TWC2019}.

\item[ii)] We propose a novel penalty-based solution \cite{QJShi_PDD_org} to tackle the IRS-dependent-QoS-constraints that are frequently encountered in various IRS design problems (e.g. feasibility-check and power minimization for (im)perfect CSI). Note that this type of constraints are generally challenging. Our solution enjoys prominent advantages over the very few ones that appear in the existing literature. For instance, the widely adopted semidefinite programming relaxation (SDR) \cite{QQWu_TWC2019,QQWu_discrete_TCom2020,MengHua_IRS_Comp,QQWu_CommLett2020} method may even fail to converge. Our method can deal with multiple IRS-dependent-QoS constraints, in contrast with \cite{Guan_CommLett2020} dealing with only one single constraint. Besides, compared to the recently proposed convex-concave penalty (CCP) method in \cite{GuiZhou_Rbst_IRS_TSP, GuiZhou_Rbst_IRS_Lett}, our solution exhibits advantages in both performance and complexity.

\item[iii)] Moreover, we also design a novel alternative direction method of multipliers (ADMM) \cite{admm_Boyd_2011} solution to deal with  IRS-dependent-QoS-constraints. Note that \cite{CM_ADMM_JunliLiangTSP2016} has utilized the standard ADMM method to handle the unit-modulus constraints in radar waveform optimization. The same idea is adopted by \cite{IRS_ADMM_QiZhang2019} for IRS design. Our proposed neADMM solution cherishes advantageous convergence and efficiency over the standard ADMM in \cite{IRS_ADMM_QiZhang2019} and the solutions in \cite{QQWu_TWC2019, MengHua_IRS_Comp, GuiZhou_Rbst_IRS_TSP, GuiZhou_Rbst_IRS_Lett}.

\item[iv)] Last but not least, we have successfully extended our proposed solutions to imperfect CSI scenarios. Specifically, we minimize the weighted sum-power with stochastic QoS constraints. This problem setting is never considered in \cite{QQWu_TWC2019, CWHuang_TWC2019, CunhuaPan_Multicell_IRS, IRS_ADMM_YCLiang2019, IRS_PWT_Ng, MengHua_IRS_Comp, QQWu_IRS_SWIPT2019, YifeiYang_OFDM_IRS, BZheng_IRS_Noma, IRS_ADMM_QiZhang2019, AI_IRS_2020, CunhuaPan_MEC_IRS, QUA_asymptotic_mmSINR, MMZhao_TTS, MMZhao_ImprctCSI, GuiZhou_Rbst_IRS_TSP, GuiZhou_Rbst_IRS_Lett, WeiXu_IRS_WCLett, WeiXu_IRS_CommLett, IRS_ChnEst_LDai, IRS_ChnEst_ZhaoruiWang, IRS_ChnEst_JieChen, IRS_practical_model} and is highly challenging since the QoS constraints should be satisfied stochastically without prior knowledge of CSI uncertainty. Via accommodating our proposed solutions into the stochastic optimization framework \cite{AnLiu_stchstc_online_2018, AnLiu_stchstc_cnstr_2019}, we successfully develop an online algorithm to tackle this problem.

\end{itemize}

\subsection{Organization of This Paper}
\label{subsec:outline}

The rest of the paper is organized as follows. Section \ref{sec:system model} introduces the system model and problem formulation. The the feasibility is discussed in Section \ref{sec:Feasibility-Check}. Section \ref{sec:max_min_sinr}, assuming perfect CSI, studies solutions to minimizing weighted sum-power. The proposed solutions are extended to imperfect CSI cases in Section \ref{sec:pwr_min_imperfect_csi}. Section Section~\ref{sec:numerical_results} presents extensive simulation results and Section \ref{sec:conclusion} concludes this paper.


\section{System Model and Problem Formulation}
\label{sec:system model}

\subsection{System Model}\label{sec-System}

As shown in Fig.\ref{fig:fig-system}, an IRS is deployed in a typical uplink MISO wireless network, where $K$ single-antenna mobile users (MUs) are communicating with a $M$-antenna access point (AP). We use $\mathcal{K}\triangleq\{1,\cdots,K\}$ to denote the set of all $K$ MUs. Suppose that the IRS has $N$ reflecting elements. The vector $\boldsymbol{h}_{\textup{d},k}\in\mathbb{C}^{M\times1}$ and $\boldsymbol{h}_{\textup{r},k}\in\mathbb{C}^{N\times1}$ denote the $k$-th MU-AP and MU-IRS link, respectively. The IRS-AP links given by $\boldsymbol{G}\in\mathbb{C}^{M\times N}$. The IRS has $N$ reflecting units, whose reflection coefficients are modeled as a complex vector $\boldsymbol{\phi}\triangleq[e^{j \theta_1}, \ldots, e^{j \theta_N}]^T$, with $\theta_n$ representing the phase-shifting of the $n$-th unit, $\forall n \in \mathcal{N}\triangleq\{1,2,\ldots, N\}$. In the following, we alternatively represent the IRS by a diagonal phase shift matrix $\boldsymbol{\Phi}\in \mathbb{C}^{N \times N}$ that is defined by $\boldsymbol{\Phi}\triangleq\textup{Diag}\big(\boldsymbol{\phi}\big)$. The effective channel $\boldsymbol{h}_k(\boldsymbol{\phi})\in \mathbb{C}^{M \times 1}$ between the $k$-th MU and the AP is given as
\begin{align}
 &\boldsymbol{h}_k(\boldsymbol{\phi})=\boldsymbol{G} \boldsymbol{\Phi}\boldsymbol{h}_{\textup{r},k} + \boldsymbol{h}_{\textup{d},k}, \ \forall k\in\mathcal{K}. \ \ \ \label{eq-def-hk}
\end{align}

The received signal at the AP can be represented by
\begin{align}
\boldsymbol{r}(\boldsymbol{\phi},\boldsymbol{q}) = \sum_{j}\sqrt{q}_j\boldsymbol{h}_j(\boldsymbol{\phi})s_j+\boldsymbol{n},  \label{eq-rcvd_sig}
\end{align}
where $s_j$ and $q_j$ are the symbol and transmission power of the $j$-th MU, respectively, and $\boldsymbol{n}\!\in\!\mathbb{C}^{M\times1}$ denotes the thermal noise. Without loss of generality, we assume $\mathbb{E}\{|s_j|\}=1$ and $\mathbb{E}\{\boldsymbol{n}\boldsymbol{n}^H\}\!=\!\sigma^2\boldsymbol{I}_M$.

Suppose AP utilizes a linear receiver $\boldsymbol{u}_k^H\in \mathbb{C}^{1\times M}$ to recover the $k$-th MU's information. Then the signal-to-interference-and-noise-ratio (SINR) for the $k$-th MU is $\gamma_k\big(\boldsymbol{\phi},\boldsymbol{q}\big) = q_k|\boldsymbol{u}_k^H \boldsymbol{h}_k(\boldsymbol{\phi})|^2/\big(\boldsymbol{u}_k^H \boldsymbol{W}_k(\boldsymbol{\phi},\boldsymbol{q} \boldsymbol){u}_k\big)$, with the matrix $\boldsymbol{W}_k(\mathbf{\phi},\boldsymbol{q})$ being defined as 
\begin{align}
\boldsymbol{W}_k( \boldsymbol{\phi},\boldsymbol{q}) = \sigma^2 \boldsymbol{I}_{M}+\sum_{i\neq k} q_i \boldsymbol{h}_i(\boldsymbol{\phi}) \boldsymbol{h}_i^H(\boldsymbol{\phi}). \label{eq-Zk-matrix}
\end{align}

When utilizing optimal filtering \cite{multiuser_MSchubert2005}, it can easily shown that the optimal SINR $\widetilde{\gamma}_k\big(\boldsymbol{\phi},\boldsymbol{q}\big)$ is given as 
\begin{align}
\widetilde{\gamma}_k\big(\boldsymbol{\phi},\boldsymbol{q}\big) = q_k\boldsymbol{h}_k^H(\boldsymbol{\phi})\big[{\boldsymbol{W}_k(\boldsymbol{\phi},\boldsymbol{q})}\big]^{-1}\boldsymbol{h}_k(\boldsymbol{\phi}). \label{eq-gamma-k-optimal}
\end{align}

\subsection{Problem Formulation} 
\label{subsec-optimizing-Mobile-IRS}

Now we clarify the problem of interest. To maintain a satisfactory QoS, each MU's SINR should be above a predefined level, i.e. $\widetilde{\gamma}_k\big(\boldsymbol{\phi},\boldsymbol{q}\big)\geq r_k$ with $r_k$ being given. At the same time, each MU has limited transmission power, i.e. $q_k\leq P_k$, $\forall k\in\mathcal{K}$. Here we aim at minimizing the total transmission power of all MUs \footnote{Although here we focus on improving MUs' battery life, as pointed out in \cite{IRS_PWT_Ng}, the power consumption owing to the IRS reflecting elements' phase-shifters can be meaningful issue to study, which will be considered in our future work}. In this paper, both perfect and imperfect CSI scenarios are considered. We first study the weighted power minimization assuming that CSI is accurate, which can be obtained via the procedure in \cite{IRS_ChnEst_LDai, IRS_ChnEst_ZhaoruiWang, IRS_ChnEst_JieChen}. Next we extends our solutions to imperfect CSI case, where both CSI itself and estimation errors are random variables.

When CSI is accurate, this problem can be formulated as 
\begin{subequations}
\label{prob:org_prob}
\begin{align}
\textup{(P1):} \quad \min_{\boldsymbol{\phi},\boldsymbol{q}}\ & \boldsymbol{\kappa}^T\boldsymbol{q} \\
\textup{s.t.} & \ \widetilde{\gamma}_k(\boldsymbol{\phi},\boldsymbol{q})\geq r_k, \forall k\in\mathcal{K}, \\
&\ q_k\leq P_k,\ \forall k\in\mathcal{K}, \\ 
&\ \ |\boldsymbol{\phi}_n|=1, \forall n\in\mathcal{N},
\end{align}
\end{subequations}
where the vector $\boldsymbol{\kappa}\triangleq[\kappa_1,\cdots,\kappa_K]^T$ denotes the weights giving the MUs different priorities.

When the CSI is imperfect, we can obtain real-time unbiased channel estimate $\widehat{\boldsymbol{\mathcal{H}}}$ of the true CSI value $\boldsymbol{\mathcal{H}}\triangleq\{\boldsymbol{G},\{\boldsymbol{h}_{\mathrm{r},k}\},\{\boldsymbol{h}_{\mathrm{d},k}\}\}$. That is $\widehat{\boldsymbol{\mathcal{H}}}=\boldsymbol{\mathcal{H}}+\boldsymbol{\Delta}$ with $\boldsymbol{\Delta}$ being the channel estimation error. At this time, the stochastic version of the power minimization problem (P1) is formulated into 
\begin{subequations}
\begin{align}
(\widetilde{\textup{P1}}): \ \min_{\boldsymbol{q},\boldsymbol{\phi}}& \ \boldsymbol{\kappa}^{T}\boldsymbol{q}, \\
& \mathbb{E}\big\{\widetilde{\gamma}_k(\boldsymbol{\phi},\boldsymbol{q},\widehat{\boldsymbol{\mathcal{H}}})\big\}\geq r_k,\ \forall k\in\mathcal{K}, \label{eq:stch_sinr_cnstr_expct_0}\\
& q_k\leq P_k,\forall k\in\mathcal{K}, \\
& |\boldsymbol{\phi}_n|=1, \forall n\in\mathcal{N}.
\end{align}
\end{subequations}
where the expectation in (\ref{eq:stch_sinr_cnstr_expct_0}) is taken over all possible realizations of $\boldsymbol{\mathcal{H}}$ and $\boldsymbol{\Delta}$ and the notation $\widetilde{\gamma}_k(\cdot,\widehat{\boldsymbol{\mathcal{H}}})$ emphasizes that $\widetilde{\gamma}_k$ depends on the channel estimates, as explained in the sequel.

Note that when CSI is inaccurate, several cases may occur: i) the true CSI $\boldsymbol{\mathcal{H}}$ is some unknown constant and $\boldsymbol{\Delta}$ is nonzero; ii) $\boldsymbol{\mathcal{H}}$ is a random variable following some unknown distribution and $\boldsymbol{\Delta}=\boldsymbol{O}$ (no estimation error); iii) both $\boldsymbol{\mathcal{H}}$ and $\boldsymbol{\Delta}$ are random variables following unknown distributions. No matter which case happens, $\mathbb{E}_{\boldsymbol{\mathcal{H}},\boldsymbol{\Delta}}\big\{\widetilde{\gamma}_k(\boldsymbol{\phi},\boldsymbol{q},\widehat{\boldsymbol{\mathcal{H}}})\big\}=\mathbb{E}_{\boldsymbol{\mathcal{H}}}\big\{\widetilde{\gamma}_k(\boldsymbol{\phi},\boldsymbol{q},\boldsymbol{\mathcal{H}})\big\}$ since the CSI estimation is unbiased. Thus the stochastic QoS constraints can be uniformly written in the form of (\ref{eq:stch_sinr_cnstr_expct_0}).

\section{Feasibility Detection} 
\label{sec:Feasibility-Check}

In this section, we first study the feasibility of the problem (P1). In fact, the significance of solving the feasibility problem is multifold. Firstly, determination of the feasibility is itself critical in scheduling reasonable QoS services for users. Second, in various optimization tasks with QoS constraints, a feasible solution is required to serve as a starting point for iterative optimization. Third, the feasibility check recurs as a key step in many complicated IRS configuration tasks, as will be shortly clear in Section \ref{sec:max_min_sinr} \& \ref{sec:pwr_min_imperfect_csi}.

\subsection{Feasibility for Sufficient Transmit Power}
\label{subsec:feasibility_pwr_cntrl}

First we study the feasibility of (P1) when transmission powers are unlimited (i.e. $P_k\rightarrow\infty$, $\forall k\in\mathcal{K}$). This problem is actually the classical feasibility characterization problem that has been extensively studied and perfectly solved \cite{power_uplink_MSchubert2004, multiuser_MSchubert2005, complete_QoS_RHunger}. When IRS is present, the feasibility region characterization is an open issue. 

Define $\boldsymbol{H}_{\mathrm{d}}\triangleq\big[\boldsymbol{h}_{\mathrm{d},1},\cdots,\boldsymbol{h}_{\mathrm{d},K}\big]$ and $\boldsymbol{H}_{\mathrm{r}}\triangleq\big[\boldsymbol{h}_{\mathrm{r},1},\cdots,\boldsymbol{h}_{\mathrm{r},K}\big]$. Then the effective channel $\boldsymbol{H}$ between the AP and all $K$ MUs is given as: 
\begin{subequations}
\begin{align}
\boldsymbol{H} = \boldsymbol{H}_{\mathrm{d}} + \boldsymbol{G}\boldsymbol{\Phi}\boldsymbol{H}_{\mathrm{r}}. 
\end{align}
\end{subequations}

Here we assume that $\boldsymbol{H}_{\mathrm{d}}$ is full rank. Note that this full rank assumption is indeed widely in the literature \cite{QQWu_TWC2019, QUA_asymptotic_mmSINR, SWIPT_JXu2014, SWIPT_QJShi2014, complete_QoS_RHunger}. In fact when channel coefficients independently follow some identical continuous distribution (e.g. Rayleigh or Ricing), the channel coefficient matrix is full rank with probability one. Here we also adopt the full rank assumption in the subsequent analysis. To simplify the transceiver design, we just assume that $M\geq K$ \cite{SWIPT_QJShi2014}. Besides, we use $\sigma_{\min}\big(\boldsymbol{X}\big)$ and $\sigma_{\max}\big(\boldsymbol{X}\big)$ to denote the minimal and maximal nonzero singular value of a matrix $\boldsymbol{X}$ respectively. Then we have the following result, which is proved in Appendix.\ref{subsec:prof_feas_cond}.

\begin{lemma}
\label{lem:feas_cond}
Assume that $\boldsymbol{H}_{\mathrm{d}}$ is of full rank and satisfies the following condition 
\begin{align}
\sigma_{\max}(\boldsymbol{G})\cdot\sigma_{\max}(\boldsymbol{H}_{\mathrm{r}})<\sigma_{\min}(\boldsymbol{H}_{\mathrm{d}}).
\end{align} 
Then for arbitrary rate constraints $\{r_k\}$, the problem (P1) with unlimited power supplies (i.e. $P_k\rightarrow\infty$,$\forall k\in\mathcal{K}$) is always feasible.
\end{lemma}
Note that Proposition 1 in \cite{QQWu_TWC2019} has obtained the sufficient condition that full rank of the composite multi-user channel can guarantee the feasibility of the MISO system assisted by IRS. Comparatively, our result in Lemma \ref{lem:feas_cond} further explores the conditions that ensure the full rank assumption adopted in \cite{QQWu_TWC2019} stands true.

\subsection{Feasibility Check for Generic Scenarios}
\label{subsec:feasibility_generic}

In this subsection, we proceed to develop a practical feasibility check method applicable to generic scenarios, which does not impose any restrictions on channel ranks or transmission powers. 

The feasibility check of (P1) can be equivalently written as the following problem: 

\begin{subequations}
\begin{align}
\textup{(P2):} \quad \sf{Find}\ & \big({\boldsymbol{\phi}},\boldsymbol{q}\big) \\
\sf{s.t.} & \ \widetilde{\gamma}_k(\boldsymbol{\phi},\boldsymbol{q})\geq r_k, \forall k\in\mathcal{K}, \\
&\ q_k\leq P_k,\ \forall k\in\mathcal{K},\\
&\ \ |\boldsymbol{\phi}_n|=1, \forall n\in\mathcal{N}.
\end{align}
\end{subequations}

To simplify the SINR constraints, we consider a closely related metric---mean square error (MSE) in the following
\begin{align}
\varepsilon_k(\boldsymbol{\phi},\boldsymbol{q},\boldsymbol{w}_k) \triangleq \mathbb{E}\Big\{\big|s_k\!-\!\boldsymbol{w}_k^H\big(\sum_{j}\sqrt{q}_j\boldsymbol{h}_js_j\!+\!\boldsymbol{n}\big)\big|^2\Big\},\label{MSE_def}
\end{align}
where $\boldsymbol{w}_k$ is the linear receiver utilized at AP to decode the $k$-th user. The expectation in (\ref{MSE_def}) is taken over the distribution of noise $\boldsymbol{n}$ and information symbols $\{s_j\}$. Then the minimal MSE $\widetilde{\varepsilon}_k(\boldsymbol{\phi},\boldsymbol{q})$ and the maximal SINR $\widetilde{\gamma}_k(\boldsymbol{\phi},\boldsymbol{q})$ is connected via the following identity:
\begin{align}
\widetilde{\varepsilon}_k(\boldsymbol{\phi},\boldsymbol{q}) = \frac{1}{1+\widetilde{\gamma}_k(\boldsymbol{\phi},\boldsymbol{q})}. \label{mse_sinr}
\end{align}

In fact, when $\big(\boldsymbol{\phi},\boldsymbol{q}\big)$ is given, the optimal linear receiver $\boldsymbol{w}_k^\star$ is obtained as the Wiener filter as follows 
\begin{align}
\boldsymbol{w}_k^{\star} = \widetilde{\boldsymbol{W}}(\boldsymbol{\phi})^{-\!1}\sqrt{q_k}\boldsymbol{h}_k(\boldsymbol{\phi}), \forall k\in\mathcal{K}, \label{eq:w_mmse_lin}
\end{align}
with $\widetilde{\boldsymbol{W}}(\boldsymbol{\phi},\boldsymbol{q})$ being
\begin{align}
\widetilde{\boldsymbol{W}}(\boldsymbol{\phi},\boldsymbol{q}) = \sigma^2\boldsymbol{I}_M+\sum_{j}q_{i}\boldsymbol{h}_j(\boldsymbol{\phi})\boldsymbol{h}_j(\boldsymbol{\phi})^H. \label{W_tilde}
\end{align}
Consequently the minimal MSE is given as $\widetilde{\varepsilon}_k(\boldsymbol{\phi},\boldsymbol{q})=1\!-\!q_k\boldsymbol{h}_k(\boldsymbol{\phi})^H\widetilde{\boldsymbol{W}}_k(\boldsymbol{\phi})^{-\!1}\boldsymbol{h}_k(\boldsymbol{\phi})$. By comparing $\widetilde{\varepsilon}_k(\boldsymbol{\phi},\boldsymbol{q})$ with (\ref{eq-gamma-k-optimal}) and employing matrix inversion lemma, (\ref{mse_sinr}) can be verified. 

Leveraging (\ref{mse_sinr}), (P2) can be equivalently written as  
\begin{subequations}
\begin{align}
\textup{(P3):} \quad \sf{Find}\ & \big({\boldsymbol{\phi}},\boldsymbol{q}\big) \\
\sf{s.t.} &\ \widetilde{\varepsilon}_k(\boldsymbol{\phi},\boldsymbol{q})\!\leq\!(1\!+\!r_k)^{-\!1}, \forall k\in\mathcal{K}, \label{eq:snr_cnstr_no_w}\\
&\ \ q_k\leq P_k, \ \forall k\in\mathcal{K}, \\
&\ \ |\boldsymbol{\phi}_n|=1, \forall n\in\mathcal{N}.
\end{align}
\end{subequations}

Noticing the relation between $\widetilde{\varepsilon}_k(\cdot)$ and $\varepsilon_k(\cdot)$, (P3) can be further transformed into the following form 
\begin{subequations}
\begin{align}
\textup{(P4):} \quad \sf{Find}\ & \big(\boldsymbol{\phi},\boldsymbol{q},\{\boldsymbol{w}_k\}\big)  \\
\sf{s.t.} &\ \varepsilon_k(\boldsymbol{\phi},\boldsymbol{q},\{\boldsymbol{w}_k\})\leq\!(1\!+\!r_k)^{-\!1}, \forall k\in\mathcal{K}, \label{eq:snr_cnstr}\\
&\ \ q_k\leq P_k, \ \forall k\in\mathcal{K}, \\
&\ \ |\boldsymbol{\phi}_n|=1, \forall n\in\mathcal{N}.
\end{align}
\end{subequations}
The equivalence between (P3) and (P4) can be seen by noticing the fact that $\boldsymbol{w}_k^\star=\mathsf{argmin}_{\boldsymbol{w}_k}\varepsilon_k(\boldsymbol{\phi},\boldsymbol{q},\boldsymbol{w}_k)$. In fact, if $\widetilde{\varepsilon}_k(\boldsymbol{\phi},\boldsymbol{q})\leq (1\!+\!r_k)^{-\!1}$, then $\varepsilon_k(\boldsymbol{\phi},\boldsymbol{q},\boldsymbol{w}_k^{\star})=\widetilde{\varepsilon}_k(\boldsymbol{\phi},\boldsymbol{q})\leq (1\!+\!r_k)^{-\!1}$, i.e. (\ref{eq:snr_cnstr}) is feasible. Conversely, if there exists $(\boldsymbol{\phi},\boldsymbol{q},\boldsymbol{w}_k)$ such that $\varepsilon_k(\boldsymbol{\phi},\boldsymbol{q},\boldsymbol{w}_k)\leq (1\!+\!r_k)^{-\!1}$, then $\varepsilon_k^{\star}(\boldsymbol{\phi},\boldsymbol{q})\leq\varepsilon_k(\boldsymbol{\phi},\boldsymbol{q},\boldsymbol{w}_k)\leq (1\!+\!r_k)^{-\!1}$, i.e. (\ref{eq:snr_cnstr_no_w}) is feasible.

Now we consider the following optimization problem  
\begin{subequations}
\begin{align}
\!\!\!\!\textup{(P5):} \ \min_{\boldsymbol{\phi},\boldsymbol{q},\{\boldsymbol{w}_k\},\alpha} \ &\alpha \\
\!\!\!\!\sf{s.t.} &\ (1\!+\!r_k)^{-\!1}\varepsilon_k(\boldsymbol{\phi},\boldsymbol{q},\boldsymbol{w}_k)\leq\alpha, \forall k\in\mathcal{K}, \\
\!\!\!\!&\ \ q_k\leq P_k, \ \forall k\in\mathcal{K}, \\
\!\!\!\!&\ \ |\boldsymbol{\phi}_n|=1, \forall n\in\mathcal{N}.
\end{align}
\end{subequations}
If the optimal value of (P5) is no greater than $1$, then (P4), namely the original problem (P1), is feasible. Otherwise the problem is infeasible. Compared to the $\widetilde{\gamma}_k(\boldsymbol{\phi},\boldsymbol{q})$ in (P2), $\varepsilon_k(\boldsymbol{\phi},\boldsymbol{q},\boldsymbol{w}_k)$ has a quadratic form and is convex with respect to each variable separately. We adopt block coordinate descent (BCD) method to solve (P5), which is elaborated subsequently.  

When $\big(\boldsymbol{\phi},\boldsymbol{q}\big)$ is fixed, the optimal $\{\boldsymbol{w}_k\}$ is given in (\ref{eq:w_mmse_lin}). When $\boldsymbol{\phi}$ and $\big\{\boldsymbol{w}_k\big\}$ are given, via introducing the intermediate variable $\boldsymbol{p}\triangleq[p_1,\cdots,p_k]$ with $p_k\triangleq\sqrt{q}_k$, $\forall k\in\mathcal{K}$, the update of $\mathbf{q}$ is equivalent to solving the following problem 
\begin{subequations}
\begin{align}
\!\!\!\!\!\!\textup{(P6):}\ \min_{\boldsymbol{p},\alpha} \ & \alpha \\
\!\!\!\!\!\!\mathsf{s.t.} &\ \sum_{k}a_{k,j}p_k^2+b_kp_k+c_k-\alpha\leq0, \forall k\in\mathcal{K}, \\
&\ \ 0\leq p_k\leq \sqrt{P_k}, \ \forall k\in\mathcal{K}, 
\end{align}
\end{subequations}
where the parameters $\{a_{k,j}\}$, $\{b_k\}$ and $\{c_k\}$ are defined as 
\begin{subequations}
\begin{align}
\!\!\!a_{k,j}\triangleq & (1\!+\!r_k)\big|\boldsymbol{w}_k^H\boldsymbol{h}_j\big|^2,\  \forall j,k\in\mathcal{K},\\
\!\!\!b_{k}\triangleq & -\!2(1\!+\!r_k)\mathsf{Re}\{\boldsymbol{w}_k^H\boldsymbol{h}_k\}, \forall k\in\mathcal{K}, \\
c_{k}\triangleq & (1\!+\!r_k)\big(\sigma^2\|\boldsymbol{w}_k\|_2^2+1\big),\ \forall k\in\mathcal{K}.
\end{align}
\end{subequations}
The problem (P6) is convex and can be efficiently solved via standard numerical sovler, like CVX \cite{CVX}. The optimal $q_k^{\star}$ can be accordingly obtained via $q_k^{\star}=p_k^{\star2}$, $\forall k\in\mathcal{K}$, with $p_k^{\star}$ being the optimal solution to (P6).

When $\{\boldsymbol{w}_k\}$ and $\boldsymbol{q}$ are given, $\boldsymbol{\phi}$ should be updated via solving the following problem:
\begin{subequations}
\begin{align}
\!\!\!\!\!\!\textup{(P7):}\ \min_{\boldsymbol{\phi},\alpha} \ &\alpha \\
\sf{s.t.} &\ \boldsymbol{\phi}^H\boldsymbol{Q}_k\boldsymbol{\phi}+2\mathsf{Re}\{\boldsymbol{q}_k^H\boldsymbol{\phi}\}+d_k\leq\alpha, \forall k\in\mathcal{K}, \label{eq:feas_phi_constr} \\
&\ \ |\boldsymbol{\phi}_n|=1, \forall n\in\mathcal{N}. \label{eq:feas_eq_constr}
\end{align}
\end{subequations}
with the parameters in (\ref{eq:feas_phi_constr}) being defined as follows:
{\small
\begin{subequations}
\begin{align}
\!\!\boldsymbol{F}_k\!\!\triangleq\!&\ \ \boldsymbol{G}\mathsf{Diag}(\boldsymbol{h}_{\mathrm{r},k}),\ \  \boldsymbol{Q}_k\triangleq(1\!+\!r_k)\sum_{j}q_j\boldsymbol{F}_j^H\boldsymbol{w}_k\boldsymbol{w}_k^H\boldsymbol{F}_j, \label{eq_cnstr_Q}\\ 
\!\!\boldsymbol{q}_k\!\!\triangleq\!& (1\!+\!r_k)\sum_{j}q_j\boldsymbol{w}_k^H\boldsymbol{h}_{\rm{d},j}\boldsymbol{F}_j^H\boldsymbol{w}_k\!-\!(1\!+\!r_k)\sqrt{q_k}\boldsymbol{F}_k^H\boldsymbol{w}_k \\
\!\!d_k\!\!\triangleq\!& (1\!+\!r_k)\big(\sum_{j}q_j\big|\boldsymbol{w}_k^H\boldsymbol{h}_{\mathrm{d},j}\big|^2\!\!-\!\!2\sqrt{q_k}\mathsf{Re}\{\boldsymbol{w}_k^H\boldsymbol{h}_{\mathrm{d},k}\}\!\!+\!\!\sigma^2\|\boldsymbol{w}_k\|_2^2\!\!+\!\!1\big).\nonumber
\end{align}
\end{subequations}
}

The major difficulty in solving (P7) lies in the coexistence of the nonlinear equality constraint (\ref{eq:feas_eq_constr}) and the convex constraints (\ref{eq:feas_phi_constr}). Here we adopt the PDD method \cite{QJShi_PDD_org}. Like the ADMM method \cite{admm_Boyd_2011}, PDD is also a kind of augmented Lagrangian (AL) method, which separates the coupled variables, via penalizing the equality constraints, and iteratively updates each variables separately. To utilize the PDD method, we introduce an auxiliary variable $\boldsymbol{\psi}$ and transform (P7) as
\begin{subequations}
\begin{align}
\!\!\!\!\!\!\textup{(P8):}\ \min_{\boldsymbol{\phi},\boldsymbol{\psi},\alpha} \ &\alpha \\
\sf{s.t.} &\ \boldsymbol{\phi}^H\boldsymbol{Q}_k\boldsymbol{\phi}+2\mathsf{Re}\{\boldsymbol{q}_k^H\boldsymbol{\phi}\}+d_k\leq\alpha, \forall k\in\mathcal{K}, \label{eq:feas_prob_rate_cnstr}\\
&\ \boldsymbol{\phi}=\boldsymbol{\psi}, \label{eq:feas_prob_xy}\\
&\ |{\boldsymbol{\psi}}_n|=1, \ \forall n\in\mathcal{N}, \label{eq:feas_prob_circle_cnstr}\\ 
&\ |\boldsymbol{\phi}_n|\leq1, \ \forall n\in\mathcal{N}.\label{eq:feas_prob_bd_cnstr}
\end{align}
\end{subequations}
Via penalizing the equality constraint, we obtain the augmented Lagrangian problem of (P8) as follows:
\begin{subequations}
\begin{align}
\!\!\!\!\!\!\textup{(P9):}\ & \min_{\boldsymbol{\phi},\boldsymbol{\psi},\boldsymbol{\lambda},\alpha}\ \alpha + \frac{1}{2\rho}\|\boldsymbol{\phi}-\boldsymbol{\psi}\|_2^2 + \mathsf{Re}\big\{\boldsymbol{\lambda}^H\big(\boldsymbol{\phi}-\boldsymbol{\psi}\big)\big\}\\
\sf{s.t.}& \ \boldsymbol{\phi}^H\boldsymbol{Q}_k\boldsymbol{\phi}+2\mathsf{Re}\{\boldsymbol{q}_k^H\boldsymbol{\phi}\}+d_k\leq\alpha, \forall k\in\mathcal{K}, \label{eq:feas_al_rate_cnstr}\\
&\ |\boldsymbol{\phi}_n|\leq1, \ \forall n\in\mathcal{N},\\
&\ |{\boldsymbol{\psi}}_n|=1, \ \forall n\in\mathcal{N}, 
\end{align}
\end{subequations}

The PDD method is a two-layer iteration, with its inner layer alternatively updating $\boldsymbol{\phi}$ and $\boldsymbol{\psi}$ and its outer layer updating the dual variables $\boldsymbol{\lambda}$ or the penalty coefficient $\rho$. The inner layer is a 2-block BCD algorithm alternatively updating $\boldsymbol{\phi}$ and $\boldsymbol{\psi}$ separately. Specifically, when $\boldsymbol{\psi}$ is fixed, the update of $\big(\boldsymbol{\phi},\alpha\big)$ is performed by solving the following problem
\begin{subequations}
\begin{align}
\!\!\!\!\!\!\textup{(P10):}\ \min_{\boldsymbol{\phi},\alpha} \ &\alpha + \frac{1}{2\rho}\|\boldsymbol{\phi}-\boldsymbol{\psi}\|_2^2 + \mathsf{Re}\big\{\boldsymbol{\lambda}^H\big(\boldsymbol{\phi}-\boldsymbol{\psi}\big)\big\},\\
\sf{s.t.} &\ \boldsymbol{\phi}^H\boldsymbol{Q}_k\boldsymbol{\phi}+2\mathsf{Re}\{\boldsymbol{q}_k^H\boldsymbol{\phi}\}+d_k\leq\alpha, \forall k\in\mathcal{K}, \label{eq:feas_al_rate_cnstr_1}\\
&\ |\boldsymbol{\phi}_n|\leq1, \ \forall n\in\mathcal{N}\label{eq:feas_al_cnstr_phi_bd}.
\end{align}
\end{subequations}
which is convex and can be numerically solved.

When $\boldsymbol{\phi}$ and $\alpha$ are given, $\boldsymbol{\psi}$ is updated by solving the following problem 
\begin{subequations}
\begin{align}
\!\!\!\!\!\!\textup{(P11):}\ \min_{\boldsymbol{\psi}} \ & \frac{1}{2\rho}\|\boldsymbol{\phi}-\boldsymbol{\psi}\|_2^2 + \mathsf{Re}\big\{\boldsymbol{\lambda}^H\big(\boldsymbol{\phi}-\boldsymbol{\psi}\big)\big\},\\
\sf{s.t.} & |{\boldsymbol{\psi}}_n|=1, \ \forall n\in\mathcal{N}.\label{eq:feas_al_rate_cnstr_2}
\end{align}
\end{subequations}

Note the fact that the quadratic term in the objective $1/(2\rho)\|\boldsymbol{\psi}\|_2^2=N/(2\rho)$ when $\psi$ has unit modulus entries. Hence the above problem (P11) is indeed reduced to 
\begin{align}
\textup{P(11)}^{\prime}:\ \max_{|\boldsymbol{\psi}|=\mathbf{1}_N}\ \mathsf{Re}\big\{\big(\boldsymbol{\phi}+\rho\boldsymbol{\lambda}\big)^H\boldsymbol{\psi}\big\}, 
\end{align}
whose maximum is attained when the elements of $\boldsymbol{\psi}$ are all aligned with those of the linear coefficient $(\rho^{\!-\!1}\boldsymbol{\phi}\!+\!\boldsymbol{\lambda})$, i.e. 
\begin{align}
\boldsymbol{\psi}^{\star}=\mathsf{exp}\Big(j\cdot\angle\big(\boldsymbol{\phi}+\rho\boldsymbol{\lambda}\big)\Big).\label{eq:psi_update}
\end{align}

The inner layer alternatively updates $\big(\boldsymbol{\phi},\alpha\big)$ and $\boldsymbol{\psi}$ until convergence is reached. For the outer layer of PDD procedure, exactly one of the two cases will occur in each iteration: 
\begin{itemize}
\item[i)] when the equality $\boldsymbol{\psi}=\boldsymbol{\phi}$ approximately holds, the Lagrangian multiplier $\boldsymbol{\lambda}$ will be updated in the same way as the standard ADMM method, which is given by
\begin{align}
\boldsymbol{\lambda}:=\boldsymbol{\lambda}+\rho^{-\!1}\big(\boldsymbol{\phi}-\boldsymbol{\psi}\big);
\end{align}
\item[ii)] alternatively, when the equality constraint $\boldsymbol{\phi}=\boldsymbol{\psi}$ is far from ``being true'', the penalty parameter $\rho^{\!-\!1}$ will be inflated, forcing the equality constraint $\boldsymbol{\phi}=\boldsymbol{\psi}$ to be approached in the subsequent iterations. 
\end{itemize}

The PDD-based algorithm to update $\boldsymbol{\phi}$ is summarized in Algorithm \ref{alg:PDD_alg} and the whole procedure to determine the feasibility (P2) is summarzied in Algorithm \ref{alg:feas_check}. Note that the $\{\eta_k\}$ (step 9 in Alg.\ref{alg:feas_check}) is a parameter sequence converging to $0$ \cite{QJShi_PDD_org} and $c$ is a positive constant smaller than $1$, whose value is typically chosen in the range of $[0.8,0.9]$. 

\begin{remark}
\label{rmrk:feas_chck_optimality}
It should be pointed out that although the optimal value $\alpha^{\star}$ of (P8) indicates the feasibility of (P1), (P8) is nonconvex in nature. Therefore the BCD-type Algorithm \ref{alg:feas_check} generally provides a sub-optimal (stationary) solution, which is an upper bound of $\alpha^{\star}$. Literally, the condition that $\alpha<1$ obtained by Algorithm~\ref{alg:feas_check} serves as a sufficient condition for the feasibility of (P1). When its output $\alpha$ is larger than 1, there exists a possibility that the true $\alpha^{\star}$ is indeed no larger than $1$. However, at this time we ``have to'' claim infeasibility since whether the optimal solution yields $\alpha^{\star}<1$ is unknown.  
\end{remark}

\begin{remark}
\label{rmrk:pdd_to_neadmm}
In Algorithm \ref{alg:feas_check}, the update of $\boldsymbol{\phi}$ (step 5) is done via invoking the PDD algorithm shown in Algorithm \ref{alg:PDD_alg}. In fact, the update of $\boldsymbol{\phi}$ (i.e. solving (P8)) can also be done using the neADMM method that will be shortly elaborated in Section \ref{sec:max_min_sinr}. 
\end{remark}

About the aforementioned PDD method, we have the following result on its convergence, which is proved in Appendix~\ref{subsec:prof_pdd_cnvg}:
\begin{theorem}
\label{thm:pdd_cnvg}
Assume that in Algorithm \ref{alg:PDD_alg}, $\big(\boldsymbol{\phi}^{(k)},\boldsymbol{\psi}^{(k)},\alpha^{(k)}\big)$ is obtained as a limit point of the sequence $\big(\boldsymbol{\phi}^{(k,t)},\boldsymbol{\psi}^{(k,t)},\alpha^{(k,t)}\big)_{t=1}^{\infty}$, $\forall k\in\{1,2,\cdots\}$. Under the assumption that $\big[\frac{1}{\rho^{(k)}}\big(\boldsymbol{\phi}^{(k)}\!\!-\!\!\boldsymbol{\psi}^{(k)}\big)\!+\!\boldsymbol{\lambda}^{(k)}\big]$ is bounded, any limit point of the sequence $\{(\boldsymbol{\phi}^{(k)},\boldsymbol{\psi}^{(k)},\alpha^{(k)})\}$ is a KKT-point of (P8).
\end{theorem}

\begin{remark}
\label{rmrk:pdd_cnvg}
Theorem \ref{thm:pdd_cnvg} is inspired by the seminal work \cite{QJShi_PDD_org}. But still, it is worth noting that our problem, compared to that in \cite{QJShi_PDD_org}, has remarkable difference in several aspects. First, our $\boldsymbol{\psi}$ has nonconvex constraint, which is more complicated than the unconstrained case (i.e. $\boldsymbol{y}$) in \cite{QJShi_PDD_org}. Second, \cite{QJShi_PDD_org} imposes the assumption that the inner-layer iterates converges to its optimality condition. Here in our case, we actually explicitly prove this is truly the fact for our specific 2-block BCD procedure. Additionally, \cite{QJShi_PDD_org} assumes the abstract regularity (Robinson's) condition in their original proof, which is a stronger assumption than ours and is generally difficulty to verify explicitly.
\end{remark}

\begin{algorithm}[!t]
\caption{PDD-based Method to Solve (P8)} \label{alg:PDD_alg}
\begin{algorithmic}[1]
\STATE initialize $\alpha^{(0)}$, $\boldsymbol{\phi}^{(0)}$, $\boldsymbol{\psi}^{(0)}$, $\boldsymbol{\lambda}^{(0)}$, $\rho^{(0)}$ and $k=1$;
\REPEAT
\STATE {\small set $\boldsymbol{\phi}^{(k\!-\!1,0)}\!:=\!\boldsymbol{\phi}^{(k\!-\!1)}$, $\boldsymbol{\psi}^{(k\!-\!1,0)}\!:=\!\boldsymbol{\psi}^{(k\!-\!1)}$, $\alpha^{(k\!-\!1,0)}\!:=\!\alpha^{(k\!-\!1)}$, $t\!=\!0$;}
\REPEAT
\STATE update $\big(\boldsymbol{\phi}^{(k\!-\!1,t+1)},\alpha^{(k\!-\!1,t+1)}\big)$ by solving (P10);
\STATE update $\boldsymbol{\psi}^{(k\!-\!1,t+1)}$ by (\ref{eq:psi_update}); $t++$;
\UNTIL{\emph{convergence}}
\STATE {\small set $\boldsymbol{\phi}^{(k)}:=\boldsymbol{\phi}^{(k\!-\!1,\infty)}$, $\boldsymbol{\psi}^{(k)}:=\boldsymbol{\psi}^{(k\!-\!1,\infty)}$, $\alpha^{(k)}:=\alpha^{(k\!-\!1,\infty)}$};
\IF{$\|\boldsymbol{\phi}^{(k)}-\boldsymbol{\psi}^{(k)}\|_{\infty}\leq\eta_k$}
\STATE $\boldsymbol{\lambda}^{(k\!+\!1)}:=\boldsymbol{\lambda}^{(k)}+\frac{1}{\rho^{(k)}}\big(\boldsymbol{\phi}^{(k)}-\boldsymbol{\psi}^{(k)}\big)$, $\rho^{(k\!+\!1)}:=\rho^{(k)}$;
\ELSE
\STATE $\boldsymbol{\lambda}^{(k\!+\!1)}:=\boldsymbol{\lambda}^{(k)}$, $\rho^{(k\!+\!1)}:=c\cdot\rho^{(k)}$;
\ENDIF
\STATE $k++$;
\UNTIL{$\|\boldsymbol{\phi}^{(t)}\!-\!\boldsymbol{\psi}^{(t)}\|_2$ is sufficiently small}
\end{algorithmic}
\end{algorithm}

{\small
\begin{algorithm}[!t]
\caption{Feasibility Check Algorithm} \label{alg:feas_check}
\begin{algorithmic}[1]
\STATE Randomly initialize $\boldsymbol{\phi}^{(0)}_n=e^{j\theta_n}$ with $\theta_n$ uniformly distributed in $[0, 2\pi)$ respectively, $\forall n\in\mathcal{N}$; randomly choose $q_k^{(0)}\in(0,P_k]$, $\forall k\in\mathcal{K}$; initialize $\{\boldsymbol{w}_k^{(0)}\}$ via (\ref{eq:w_mmse_lin}), set 
$\alpha^{(0)}:=\max_{k}\varepsilon_k(\boldsymbol{\phi}^{(0)},\boldsymbol{q}_k^{(0)}, \boldsymbol{w}_k^{(0)})$, $t:=0$;
\REPEAT
\STATE update $\boldsymbol{w}_k^{(t+1)}$ via (\ref{eq:w_mmse_lin});
\STATE update $\alpha^{(t+1)}$ and $\boldsymbol{q}^{(t+1)}$ by solving (P6);
\STATE update $\alpha^{(t+1)}$ and $\boldsymbol{\phi}^{(t+1)}$ by invoking Alg.\ref{alg:PDD_alg};
\STATE $t++$;
\UNTIL{\emph{convergence or} $\alpha^{(t)}\leq1$}
\IF{$\alpha^{(t)}<=1$}
\STATE claim {\sf{Feasible}}, $\big(\boldsymbol{\phi}^{(t)},\boldsymbol{q^{(t)}}\big)$ is a feasible solution;
\ELSE
\STATE claim {\sf{Infeasible}};
\ENDIF
\end{algorithmic}
\end{algorithm}
}

\section{Power Minimization For Perfect CSI}
\label{sec:max_min_sinr}

In this section we study the power minimization problem (P1) assuming that CSI is perfectly known. Based on the previous discussion, we just assume that (P1) is feasible. 

By utilizing the \emph{multidimentional quadratic transform} proposed in \cite{KMShen_2018}, we first rewrite the receive SINR in (\ref{eq-gamma-k-optimal}) as
{\small
\begin{align}
\widetilde{\gamma}_{k}(\boldsymbol{\phi},\boldsymbol{q})&=\big(\sqrt{q_k}\boldsymbol{h}_k\big)^H\boldsymbol{W}_k^{-1}\big(\sqrt{q_k}\boldsymbol{h}_k\big) \nonumber \\
&=\max_{\boldsymbol{y}_k}\Big\{\ 2\mathrm{Re}\big\{\sqrt{q_k}\boldsymbol{y}_k^H\boldsymbol{h}_k\big\}-\boldsymbol{y}_k^H\boldsymbol{W}_k\boldsymbol{y}_k\Big\}. \label{eq:y_max}
\end{align}
}
The above identity can be readily verified by noticing that the objective in the maximization in (\ref{eq:y_max}) is concave with respect to $\boldsymbol{y}_k$ and its maximum is obtained when 
\begin{shrinkeq}{-1ex}
\begin{align}
\boldsymbol{y}_k^{\star} = \sqrt{q}_k\boldsymbol{W}_k^{-1}\boldsymbol{h}_k. \label{eq:y_update}
\end{align}
\end{shrinkeq}

Therefore the power minimization problem in (P1) can be equivalently written as
\begin{shrinkeq}{-1ex}
\begin{subequations}
\begin{align}
\!\!\!\!\!\!\!\!\textup{(P12)}&\min_{\{\boldsymbol{y}_k\},\boldsymbol{q},\boldsymbol{\phi}} \boldsymbol{\kappa}^{T}\boldsymbol{q}, \\
\!\!\!\!\!\!\!\!& 2\mathsf{Re}\big\{\boldsymbol{y}_k^H\boldsymbol{h}_k\big\}-\boldsymbol{y}_k^H\boldsymbol{W}_k\boldsymbol{y}_k \geq r_k, \forall k\in\mathcal{K}, \\
\!\!\!\!\!\!\!\!& q_k\leq P_k,\forall k\in\mathcal{K}, \\
\!\!\!\!\!\!\!\!& |\boldsymbol{\phi}_n|=1, \forall n\in\mathcal{N}.
\end{align}
\end{subequations}
\end{shrinkeq}

To solve the above problem, we still apply BCD method. The alternative updating steps are given as follows. When $\big(\boldsymbol{\phi},\boldsymbol{q}\big)$ are given, $\boldsymbol{y}_k$ can be obtained in a closed form according to (\ref{eq:y_update}). When $\big(\boldsymbol{\phi},\{\boldsymbol{y}_k\}\big)$ are fixed, we introduce intermediate variables $\boldsymbol{p}\triangleq[p_1,\cdots,p_K]$ with $p_k\triangleq\sqrt{q_k}$. Then the update of $\boldsymbol{q}$ is equivalent to solving 
\begin{shrinkeq}{-1ex}
\begin{subequations}
\begin{align}
\!\!\!\!\!\!\textup{(P13):\ }& \min_{\boldsymbol{p}\geq0} \quad \sum_k \kappa_kp_k^2 \\
\textup{s.t.} & \sum_{j\neq k}\tilde{a}_{k,j}p_j^2 + \tilde{b}_kp_k + \tilde{c}_k + r_k\leq0, \ \ \forall k\in\mathcal{K}, \\
& 0\leq p_k \leq \sqrt{P}_k, \forall k\in\mathcal{K},
\end{align}
\end{subequations}
\end{shrinkeq}
with the parameters $\{\tilde{a}_{k,j}\}$, $\{\tilde{b}_k\}$ and $\{\tilde{c}_k\}$ defined as follows:
\begin{align}
\!\!\!\!\tilde{a}_{k,j}\triangleq|\boldsymbol{y}_k^H\boldsymbol{h}_j|^2,\ \ 
\tilde{b}_{k}\triangleq-2\mathsf{Re}\{\boldsymbol{y}_k^H\boldsymbol{h}_k\},\ \ 
\tilde{c}_{k}\triangleq\sigma^2\|\boldsymbol{y}_k\|_2^2.
\end{align}
The corresponding update of $q_k$ is given by $q_k=p_k^{\star2}$ with $\{p_k^\star\}$ being the optimal solution to (P13).

When $\big(\{\boldsymbol{y}_k\},\boldsymbol{q}\big)$ are fixed, the update of $\boldsymbol{\phi}$ can rewritten as 
\begin{subequations}
\begin{align}
\!\!\!\!\!\!\textup{(P14):\ }& \mathsf{Find}\ \ \boldsymbol{\phi} \\
\!\!\!\!\!\!\textup{s.t.\ } & \boldsymbol{\phi}^H\boldsymbol{S}_k\boldsymbol{\phi}+2\mathsf{Re}\big\{\boldsymbol{s}_k^H\boldsymbol{\phi}\big\}+s_k\leq0, \forall k\in\mathcal{K}, \label{eq:pwr_phi_cnstr}\\ 
\!\!\!\!& |\boldsymbol{\phi}_n| = 1,\forall n\in\mathcal{N}. 
\end{align}
\end{subequations}
with the parameters in (\ref{eq:pwr_phi_cnstr}) defined as follows:
{\small
\begin{align}
\!\!\!\!\!\boldsymbol{S}_k & \triangleq\sum_{j\neq k}q_j\boldsymbol{F}_{\mathrm{r},j}^H\boldsymbol{y}_k\boldsymbol{y}_k^H\boldsymbol{F}_{\mathrm{r},j}, \forall k \in \mathcal{K}, \\
\!\!\!\!\!\boldsymbol{s}_k & \triangleq\sum_{j\neq k}q_j\big(\boldsymbol{h}_{\mathrm{d},j}^H\boldsymbol{y}_k\big)^*\boldsymbol{F}_{\mathrm{r},j}^H\boldsymbol{y}_k-\sqrt{q_k}\boldsymbol{F}_{\mathrm{r},k}^H\boldsymbol{y}_k, \forall k \in \mathcal{K}, \nonumber\\
\!\!\!\!\!s_k & \triangleq \sigma^2\|\boldsymbol{y}_k\|_2^2\!+\!\sum_{j\neq k}q_j|\boldsymbol{h}_{\mathrm{d},j}^H\boldsymbol{y}_k|^2\!-\!2\mathsf{Re}\big\{\sqrt{q_k}\boldsymbol{y}_k^H\boldsymbol{h}_{\mathrm{d},k}\big\}\!+\!r_k.\nonumber
\end{align}
}

Note that solving (P14) becomes somewhat tricky since the objective of (P14) is independent of $\boldsymbol{\phi}$. To wisely configure IRS elements' phase shifting, we turn to solve the following \emph{feasibility} problem:
\begin{subequations}
\begin{align}
\!\!\!\!\!\!(\widetilde{\textup{P14}}):\ & \min_{\boldsymbol{\phi},\beta} \quad \beta \\
\!\!\!\!\!\!\textup{s.t.\ } & \boldsymbol{\phi}^H\boldsymbol{S}_k\boldsymbol{\phi}+2\mathsf{Re}\big\{\boldsymbol{s}_k^H\boldsymbol{\phi}\big\}+s_k\leq \beta, \forall k\in\mathcal{K}, \\ 
\!\!\!\!& |\boldsymbol{\phi}_n| = 1,\forall n\in\mathcal{N}. 
\end{align}
\end{subequations}

The update of $\boldsymbol{\phi}$ by solving $(\widetilde{\textup{P14}})$ ensures that the newly obtained $(\boldsymbol{q},\boldsymbol{\phi})$ is still feasible to (P1). In fact, since the previous iterates are all feasible (recall that we start the entire BCD procedure from a feasible solution), the $\beta$ yielded by the current $\boldsymbol{\phi}$ should be non-positive and therefore the optimal $\beta$ associated with the updated $\boldsymbol{\phi}$ should also be non-positive. Intuitively, by solving $(\widetilde{\textup{P14}})$, we actually push $\boldsymbol{\phi}$ away from the boundaries of the feasible region towards a ``more feasible'' solution, which provides larger SINR margin to further reduce the sum-power in the subsequent iterations.

To attack the problem $(\widetilde{\textup{P14}})$, one option is to utilize the PDD method again, as discussed previously. Here we employ the novel framework proposed by the very recent work \cite{neADMM2019}, namely the nonlinear equality constrained alternating direction method of multipliers (neADMM), to handle the above problem. Consider the following problem \cite{neADMM2019} 
\begin{subequations}
\label{eq:neADMM_formulatoin}
\begin{align}
& \min_{\boldsymbol{x}_1,\boldsymbol{x}_2}\ F_1(\boldsymbol{x}_1) + F_2(\boldsymbol{x}_2), \\
& s.t.\ \boldsymbol{f}_1(\boldsymbol{x}_1) + \boldsymbol{f}_2(\boldsymbol{x}_2)=\mathbf{0},
\end{align}
\end{subequations}
where $F_1(\boldsymbol{x}_1)$ and $F_2(\boldsymbol{x}_1)$ are convex functions and $\boldsymbol{f}_1(\boldsymbol{x}_1)$ and $\boldsymbol{f}_2(\boldsymbol{x}_2)$ are high-dimension nonlinear functions. \cite{neADMM2019} extends the conventional ADMM \cite{admm_Boyd_2011}, which is originally proposed to tackle linear-equality-constrained problems, to the ones constrained by nonlinear equalities. Specifically, to solve (\ref{eq:neADMM_formulatoin}), we optimize its AL function    
\begin{align}
\!\!\!\!L(\boldsymbol{x}_1,\boldsymbol{x}_2,\boldsymbol{z})\!&=\!F_1(\boldsymbol{x}_1)\!+\!F_2(\boldsymbol{x}_2)\!+\!\boldsymbol{z}^T\big(\boldsymbol{f}_1(\boldsymbol{x}_1)\!+\!\boldsymbol{f}_2(\boldsymbol{x}_2)\big) \nonumber\\
&\qquad + \rho/2\|\boldsymbol{f}_1(\boldsymbol{x}_1)\!+\!\boldsymbol{f}_2(\boldsymbol{x}_2)\|_2^2
\end{align}
in an ADMM manner, as shown in Algorithm~\ref{alg:neADMM_framework}.

{\small
\begin{algorithm}[!t]
\caption{neADMM Framework} \label{alg:neADMM_framework}
\begin{algorithmic}[1]
\STATE Initialize $\boldsymbol{x}_1$, $\boldsymbol{x}_2$, $\boldsymbol{z}$, $k=0$;
\REPEAT 
\STATE Update $\boldsymbol{x}_1^{(k\!+\!1)}:=\mathsf{argmin}_{\boldsymbol{x}_1}L(\boldsymbol{x}_1,\boldsymbol{x}_2^{(k)},\boldsymbol{z}^{(k)})$;
\STATE Update $\boldsymbol{x}_2^{(k\!+\!1)}:=\mathsf{argmin}_{\boldsymbol{x}_2}L(\boldsymbol{x}_1^{(k\!+\!1)},\boldsymbol{x}_2,\boldsymbol{z}^{(k)})$;
\STATE Update $\boldsymbol{z}^{(k\!+\!1)}\!\!:=\!\boldsymbol{z}^{(k)}\!\!+\!\!\rho\big(\boldsymbol{f}_1(\boldsymbol{x}_1^{(k\!+\!1)})\!\!+\!\!\boldsymbol{f}_2(\boldsymbol{x}_2^{(k\!+\!1)})\big)$, $\!k\!+\!+\!$;
\UNTIL{$\|\boldsymbol{z}^{(k\!+\!1)}\!-\!\boldsymbol{z}^{(k)}\|_2$ is sufficiently small}
\end{algorithmic}
\end{algorithm}
}

To employ the neADMM framework, we first introduce the auxiliary variable $\boldsymbol{\psi}$ to decouple constraints as follows
\begin{subequations}
\begin{align}
\!\!\!\!\!\!\!\!\!\!\!\!\textup{(P15):\ }\min_{\beta,\boldsymbol{\phi},\boldsymbol{\psi}}& \beta \\
\!\!\!\!\!\!\!\!\!\!\!\!\textup{s.t.\ } & \boldsymbol{\phi}^H\boldsymbol{S}_k\boldsymbol{\phi}+2\mathsf{Re}\big\{\boldsymbol{s}_k^H\boldsymbol{\phi}\big\}+s_k\leq\beta, \forall k\in\mathcal{K}, \label{eq:maxmin_prob_ineq_cnstr} \\
\!\!\!\!\!\!\!\!\!\!\!\!& \boldsymbol{\phi} = \boldsymbol{\psi}, \label{eq:maxmin_prob_equ_cnstr} \\
\!\!\!\!\!\!\!\!\!\!\!\!& |\boldsymbol{\psi}| = \mathbf{1}_N. \label{eq:maxmin_prob_modulus_cnstr}
\end{align}
\end{subequations}

To accommodate (P15) into the neADMM framework, we identify the variables $\boldsymbol{x}_1$ and $\boldsymbol{x}_2$ in (\ref{eq:neADMM_formulatoin}) as $\boldsymbol{x}_1\triangleq[\beta, \boldsymbol{\phi}^T]^T$ and $\boldsymbol{x}_2\triangleq\boldsymbol{\psi}$, respectively. Besides, the objective $F_1(\boldsymbol{x}_1)$ and $F_2(\boldsymbol{x}_2)$ and the constraint $\boldsymbol{f}_1(\boldsymbol{x}_1)\in\mathbb{R}^{2N}$ and $\boldsymbol{f}_2(\boldsymbol{x}_2)\in\mathbb{R}^{2N}$ in (\ref{eq:neADMM_formulatoin}) are given as follows:
\begin{subequations}
\begin{align}
\!\!F_1(\beta,\boldsymbol{\phi})&\triangleq \beta+\mathbb{I}_{\mathcal{X}}(\beta,\boldsymbol{\phi}), \quad F_2(\boldsymbol{\psi})\triangleq 0, \\
\!\!\boldsymbol{f}_1(\beta,\boldsymbol{\phi})&=[-\boldsymbol{\phi}^T,\mathbf{0}_N^T]^T,\ 
\boldsymbol{f}_2(\boldsymbol{\psi})=[\boldsymbol{\psi}^T,|\boldsymbol{\psi}|^T\!-\!\mathbf{1}_N^T]^T,
\end{align}
\end{subequations}
where $\mathbb{I}_{\mathcal{X}}(\cdot)$ is the indicator function of the set $\mathcal{X}$ and $\mathcal{X}$ is the feasible region defined by the bunch of inequalities presented in (\ref{eq:maxmin_prob_ineq_cnstr}). Then the AL function of (P15) is defined as 
\begin{align} 
\!\!\!L(\beta,\boldsymbol{\phi},\boldsymbol{\psi},\boldsymbol{z}_1,\boldsymbol{z}_2)
&\triangleq\beta\!+\!\mathbb{I}_{\mathcal{X}}(\beta,\boldsymbol{\phi})\!+\!\rho/2\big\||\boldsymbol{\psi}|\!-\!\mathbf{1}_N+\rho^{\!-\!1}\boldsymbol{z}_1\big\|_2^2 \nonumber \\
& \qquad + \rho/2\big\|\boldsymbol{\psi}-\boldsymbol{\phi}+\rho^{\!-\!1}\boldsymbol{z}_2\big\|_2^2,
\end{align}
with $\boldsymbol{z}_1$ and $\boldsymbol{z}_2$ being the Lagrangian multipliers associated with the constraints (\ref{eq:maxmin_prob_modulus_cnstr}) and (\ref{eq:maxmin_prob_equ_cnstr}), respectively. 

Following the previous exposition, we proceed to elaborate the update steps in minimizing $L(\cdot)$. The update of $\big(\beta, \boldsymbol{\phi}\big)$ can be done by solving the subsequent convex problem
\begin{subequations}
\begin{align}
\!\!\!\!\!\!\textup{(P16):\ } \min_{\beta, \boldsymbol{\phi}} &\ \beta + \frac{\rho}{2}\big\|\boldsymbol{\psi}-\boldsymbol{\phi}+\rho^{-1}\boldsymbol{z}_2\big\|_2^2 \\
\textup{s.t.} & \ \boldsymbol{\phi}^H\boldsymbol{S}_k\boldsymbol{\phi}+2\mathsf{Re}\big\{\boldsymbol{s}_k^H\boldsymbol{\phi}\big\}+s_k\leq\beta, \forall k\in\mathcal{K}.
\end{align}
\end{subequations}

To update $\boldsymbol{\psi}$, we need to solve the following problem
\begin{align}
\!\!\!\!\textup{(P17):}\ & \min_{\boldsymbol{\psi}} \big\|\boldsymbol{\psi}\!-\!\boldsymbol{\phi}\!+\!\rho^{\!-\!1}\boldsymbol{z}_2\big\|_2^2 + \big\| |\boldsymbol{\psi}|\!-\!\mathbf{1}_N\!+\!\rho^{\!-\!1}\boldsymbol{z}_1\big\|_2^2. 
\end{align}
The problem (P17) is a non-convex problem. Fortunately, it has an analytic solution as stated in the following lemma, whose proof can be found in Appendix \ref{subsec:prof_psi_solution}

\begin{lemma}
\label{lem:psi_solution}
The optimal solution $\boldsymbol{\psi}^{\star}$ to $(P17)$, with $\psi_i^{\star}$ being its $i$-th entry, is given as follows
\begin{align}
\psi_i^{\star} = \frac{s_i}{2-t_i/|\psi_i^{\star}|},\ \ \forall i\in\mathcal{N}, 
\end{align}
where $s_i\triangleq\phi_i-\rho^{\!-\!1}z_{2,i}, \ t_i\triangleq\mathsf{Re}\{1-\rho^{-1}z_{1,i}\}$ and the value of $|\psi_i^{\star}|$ can be determined by solving the following equation 
\begin{align}
\big|2|\psi_i^{\star}|-t_i\big|=|s_i|.
\end{align}
\end{lemma}

When both $(\beta,\boldsymbol{\phi})$ and $\boldsymbol{\psi}$ are fixed, $\boldsymbol{z}_1$ and $\boldsymbol{z}_2$ are updated in a gradient descent manner specified as follows 
\begin{subequations}
\label{eq:z_update}
\begin{align}
\boldsymbol{z}_1^{(t\!+\!1)} &= \boldsymbol{z}_1^{(t)}+\rho\big(\big|\boldsymbol{\psi}^{(t\!+\!1)}\big|-\mathbf{1}_N\big) \\ 
\boldsymbol{z}_2^{(t\!+\!1)} &= \boldsymbol{z}_2^{(t)}+\rho\big(\boldsymbol{\psi}^{(t\!+\!1)}-\boldsymbol{\phi}^{(t\!+\!1)}\big).
\end{align}
\end{subequations}

The entire neADMM-based solution to minimize sum-power is summarized in Algorithm~\ref{alg:opt-overall}. 
\begin{remark}
\label{rmrk:neadmm_cnvg}
A convergence proof is given in \cite{neADMM2019}, which is based on the assumption of existence of saddle point(s). However, this assumption generally does not hold true for non-convex problems. Therefore the optimality of neADMM's solutions in solving $(\widetilde{\textit{P14}})$ is still an open issue, although it always works well in practice. 
\end{remark}

\begin{remark}
\label{rmrk:neadmm_to_pdd}
In Algorithm \ref{alg:opt-overall}, the update of $\boldsymbol{\phi}$ (step 8-10) is done via neADMM algorithm following the steps in Algorithm \ref{alg:neADMM_framework}. Indeed, the neADMM algorithm can be substituted by the PDD algorithm. The performance of both the PDD-based and neADMM-based solutions' are presented and compared in Section \ref{sec:numerical_results}.
\end{remark}

{\small
\begin{algorithm}[!t]
\caption{Solving (P1)} \label{alg:opt-overall}
\begin{algorithmic}[1]
\STATE Invoke Alg.~\ref{alg:feas_check} to obtain a feasible $(\boldsymbol{\phi}^{(0)}, \boldsymbol{q}^{(0)})$; $t=0$;
\STATE Initialize $\{\boldsymbol{y}_k^{(0)}\}$ via (\ref{eq:y_update});
\REPEAT 
\STATE Update $\{\boldsymbol{y}_k^{(t\!+\!1)}\}$ by (\ref{eq:y_update});
\STATE Update $\big(\boldsymbol{q}^{(t\!+\!1)}\big)$ solving (P12); 
\STATE $u=0$, set $\boldsymbol{\phi}^{(t,0)}=\boldsymbol{\phi}^{(t)}$;
\REPEAT
\STATE update $\big(\boldsymbol{\phi}^{(t,u)},\boldsymbol{\beta}^{(u)}\big)$ by solving (P16);
\STATE update $\boldsymbol{\psi}^{(u)}$ by Lemma.\ref{lem:psi_solution};
\STATE update $\boldsymbol{z}_1^{(u)}$ and $\boldsymbol{z}_2^{(u)}$ by (\ref{eq:z_update});\ $u++$;
\UNTIL{$\|\boldsymbol{\psi}^{(t,u)}\!\!-\!\boldsymbol{\phi}^{(u)}\|_{2}$ \& $\|\boldsymbol{\phi}^{(u)}\!-\!\mathbf{1}_N\|_{2}$ small}
\STATE$\boldsymbol{\phi}^{(t\!+\!1)}\!:=\!\boldsymbol{\phi}^{(t,\infty)}$, \ $t\!+\!+$; 
\UNTIL{\textit{convergence}}
\end{algorithmic}
\end{algorithm}
}

\section{Power Minimization for Imperfect CSI}
\label{sec:pwr_min_imperfect_csi}



In this section, we proceed to study power minimization when CSI is imperfect. Before elaborating our solutions, we need to point it out that one strategy to combat CSI uncertainty is robust design \cite{GuiZhou_Rbst_IRS_TSP}. Robust beamforming mainly falls into two categories: i) stochastic model based method and ii) bounded ellipsoid model based method. The former assumes that the CSI follows Gaussian distribution and the latter assumes that all CSI errors live in a bounded ellipsoid. Robust design has some inherent shortcomings. First, it heavily depends on the prior knowledge of the CSI model (i.e. Gaussian distribution statistics or ellipsoid bounds), which is generally unknown in practice. Moreover, robust optimization is in nature a kind of conservative design. It shrinks the feasible region to ensure the ``worst-case'' performance and consequently degrades the stochastic average performance, which is usually the genuine target we pursue.

In contrast to robust design, the online stochastic framework in \cite{AnLiu_stchstc_online_2018, AnLiu_stchstc_cnstr_2019} does not require any prior knowledge of the CSI model or channel estimation errors and can effectively tackle the expectation in the problem. To simplify the following exposition, we assume that in each time \emph{slot} we can obtain a new sample of channel estimate. We rewrite the problem ($\widetilde{\textup{P1}}$) in the sequel
{\small
\begin{shrinkeq}{-1ex}
\begin{subequations}
\label{prb:stch_pwr_cnstr_expct}
\begin{align}
\textup{(P18):\ }\min_{\boldsymbol{q},\boldsymbol{\phi}}& \ \boldsymbol{\kappa}^{T}\boldsymbol{q}, \\
& \mathbb{E}\big\{\widetilde{\gamma}_k(\boldsymbol{\phi},\boldsymbol{q},\widehat{\boldsymbol{\mathcal{H}}})\big\}\geq r_k,\ \forall k\in\mathcal{K}, \label{eq:stch_sinr_cnstr_expct}\\
& q_k\leq P_k,\forall k\in\mathcal{K}, \\
& |\boldsymbol{\phi}_n|=1, \forall n\in\mathcal{N}.
\end{align}
\end{subequations}
\end{shrinkeq}
}

Note the expectation in (\ref{eq:stch_sinr_cnstr_expct}) cannot be transformed into an explicit form. To overcome this difficulty, following the strategy in \cite{AnLiu_stchstc_online_2018, AnLiu_stchstc_cnstr_2019}, we replace it by a combination of successive convex approximations of historical SINRs and approach to its true value via learning from the consistently incoming fresh CSI estimates in a online manner. In the $t$-th time slot, a potentially inaccurate CSI estimate $\widehat{\boldsymbol{\mathcal{H}}}^{(t)}$ is obtained. Then a surrogate function approximating the expectation in (\ref{eq:stch_sinr_cnstr_expct}) is constructed as follows:
\begin{align}
\label{eq:sinr_surrogate}
\widehat{\gamma}_{k}^{(t)}(\boldsymbol{\phi},\boldsymbol{q}) & = (1\!-\!\rho^{(t)})\widehat{\gamma}_{k}^{(t\!-\!1)}(\boldsymbol{\phi},\boldsymbol{q}) \\
& \qquad\qquad + \rho^{(t)}\overline{\gamma}_k\big(\boldsymbol{\phi},\boldsymbol{q},\boldsymbol{\phi}^{(t\!-\!1)},\boldsymbol{q}^{(t\!-\!1)},\widehat{\boldsymbol{\mathcal{H}}}^{(t)}\big)\nonumber
\end{align}
where the function $\overline{\gamma}_k(\boldsymbol{\phi},\boldsymbol{q},\boldsymbol{\phi}^{(t\!-\!1)},\boldsymbol{q}^{(t\!-\!1)},\widehat{\boldsymbol{\mathcal{H}}}^{(t)})$ is given as
\begin{align}
\label{eq:sinr_surrogate_incrs}
\!\!\!\!\!\!&\overline{\gamma}_k\big(\boldsymbol{\phi},\boldsymbol{q},\boldsymbol{\phi}^{(t\!-\!1)},\boldsymbol{q}^{(t\!-\!1)},\widehat{\boldsymbol{\mathcal{H}}}^{(t)}\big)\triangleq 
\widetilde{\gamma}_{k}(\boldsymbol{\phi}^{(t\!-\!1)},\boldsymbol{q}^{(t\!-\!1)},\widehat{\boldsymbol{\mathcal{H}}}^{(t)})\! \nonumber\\
&\quad\qquad + 2\mathsf{Re}\Big\{\Big(\frac{\partial \widetilde{\gamma}_k}{\partial \boldsymbol{\phi}}\Big)^H\!\big(\boldsymbol{\phi}\!-\!\boldsymbol{\phi}^{(t\!-\!1)}\big)\Big\} + \Big(\frac{\partial \widetilde{\gamma}_k}{\partial \boldsymbol{q}}\Big)^T\!\big(\boldsymbol{q}-\boldsymbol{q}^{(t\!-\!1)}\big) \nonumber\\
&\qquad\qquad - \frac{\tau_k}{2}\|\boldsymbol{\phi}\!-\!\boldsymbol{\phi}^{(t\!-\!1)}\|_2^2 - \frac{\tau_k}{2}\|\boldsymbol{q}-\boldsymbol{q}^{(t\!-\!1)}\|_2^2,
\end{align}
with $\tau_k$ being any positive constant, $\forall k\in\mathcal{K}$. Besides, the objective in (P18) should also be replaced via the a surrogate function $f_0(\boldsymbol{q})$ having a proximal term to ensure convergence, which is defined as follows:
\begin{align}
\label{eq:obj_surrogate}
\!\!\!f_0^{(t)}(\boldsymbol{q})\!=\!(1\!\!-\!\!\rho^{(t)})f_0^{(t\!-\!1)}\!(\boldsymbol{q})\!\!+\!\!\rho^{(t)}\Big(\!\boldsymbol{\kappa}^T\boldsymbol{q}\!+\!\!\frac{\tau_0}{2}\|\boldsymbol{q}\!-\!\boldsymbol{q}^{(t\!-\!1)}\|_2^2\Big)
\end{align}

Following the paradigm in \cite{AnLiu_stchstc_cnstr_2019}, the stochastic problem (P18) is solved in an online fashion via repeatedly solving the surrogate problem, which approximates the original problem's intractable expectations using the surrogate functions that are constantly rectified by real-time samples. Specifically, each slot the surrogate functions are updated by a newly obtained channel estimate, we need first check the feasibility of (P18)'s surrogate problem, shown as below
\begin{shrinkeq}{-1ex}  
\begin{subequations}
\label{prb:stchs_feas}
\begin{align}
\textup{(P19):}\ \min_{\boldsymbol{q},\boldsymbol{\phi}}\ & \beta \\
& -\widehat{\gamma}_k^{(t)}\big(\boldsymbol{q},\boldsymbol{\phi}\big)+\!r_k\leq\beta,\ k\in\mathcal{K}, \label{eq:stch_sinr_cnstr}\\
& q_k\leq P_k,\forall k\in\mathcal{K}, \\
& |\boldsymbol{\phi}_n|=1, \forall n\in\mathcal{N}.
\end{align}
\end{subequations}
\end{shrinkeq}

Due to the fluctuation of the CSI, when a new slot's $\widehat{\boldsymbol{\mathcal{H}}}$ arrives, the problem (P19) may be feasible or not. If it is infeasible (i.e. the optimal $\beta^\star$ is larger than $0$), the solution to (P19) is used to update the $(\boldsymbol{q},\boldsymbol{\phi})$. Otherwise, we proceed to solve the surrogate power minimization problem in the sequel 
\begin{shrinkeq}{-1ex}
\begin{subequations}
\label{prb:stchs_pwr_min}
\begin{align}
\textup{(P20):}\ \min_{\boldsymbol{q},\boldsymbol{\phi}}\ & f_0^{(t)}(\boldsymbol{q}) \\
& -\widehat{\gamma}_k^{(t)}\big(\boldsymbol{q},\boldsymbol{\phi}\big)+\!r_k\leq0, k\in\mathcal{K}, \label{eq:stch_sinr_cnstr_2}\\
& q_k\leq P_k,\forall k\in\mathcal{K}, \\
& |\boldsymbol{\phi}_n|=1, \forall n\in\mathcal{N},
\end{align}
\end{subequations}
\end{shrinkeq}

Assume that the new $\widetilde{\boldsymbol{\phi}}$ and $\widetilde{\boldsymbol{q}}$ are obtained by solving (P19) or (P20) (when (P19) is feasible), then $\boldsymbol{\phi}, \boldsymbol{q}$ are updated as follows:
\begin{subequations}
\label{prb:stchs_phi_q_update}
\begin{align}
\widetilde{\boldsymbol{\theta}}&\triangleq\angle{\widetilde{\boldsymbol{\phi}}},\ \ \  \boldsymbol{\theta}^{(t\!-\!1)}\triangleq\angle{\boldsymbol{\phi}^{(t\!-\!1)}}, \\
\boldsymbol{\theta}^{(t)}&:=(1\!-\!\rho^{(t)})\boldsymbol{\theta}^{(t\!-\!1)}+\rho^{(t)}\widetilde{\boldsymbol{\theta}}, \\
\boldsymbol{\phi}^{(t)}&:=\mathsf{exp}(j\cdot\boldsymbol{\theta}^{(t)}), \\
\boldsymbol{q}^{(t)}&:=(1\!-\!\rho^{(t)})\boldsymbol{q}^{(t\!-\!1)}+\rho^{(t)}\widetilde{\boldsymbol{q}}. 
\end{align}
\end{subequations}

The whole procedure of the online minimization of sum-power is summarized in the Algorithm~\ref{alg:stchs-pwr_min}. The underlying intuition of Algorithm~\ref{alg:stchs-pwr_min} is straightforward. When (P18) is estimated infeasible via solving (P19), the variables move towards the feasible region in a stochastic sense, which is reminiscent of the feasibility update in (P14) for perfect CSI case. When the surrogate function is approximately feasible, we then turn to minimize the sum-power. 

Next we discuss how to tackle the two sub-problems (P19) and (P20). By the iterative definition of the surrogate functions in (\ref{eq:sinr_surrogate}-\ref{eq:obj_surrogate}), it can be readily seen that the constraint (\ref{eq:stch_sinr_cnstr}) or (\ref{eq:stch_sinr_cnstr_2}) is convex quadratic with respect to $(\boldsymbol{q},\boldsymbol{\phi})$ jointly $\forall k \in \{0\}\cup\mathcal{L}$ and any $t$. The only non-convexity of (P19) and (P20) stems from the constant modulus constraint $|\boldsymbol{\phi}|=\boldsymbol{1}_N$. Obviously we can still use the PDD or neADMM method to solve these problems. Since the application of PDD or neADMM method to (P19) and (P20) is rather straightforward, we omit the details for brevity.

{\small
\begin{algorithm}[!t]
\caption{Power Min. w. Stchs. SINR Cnstr.} \label{alg:stchs-pwr_min}
\begin{algorithmic}[1]
\STATE Initialize $f_k^{(0)}(\boldsymbol{q},\boldsymbol{\phi}):=0$, $\forall k\in\{0\}\cup\mathcal{L}$; $t:=1$;
\STATE Randomly generate feasible $\boldsymbol{q}^{(0)}$ and $\boldsymbol{\phi}^{(0)}$. 
\REPEAT 
\STATE solve (P19) to obtain $(\widetilde{\beta},\widetilde{\boldsymbol{q}},\widetilde{\boldsymbol{\phi}})$;
\IF{$\widetilde{\beta}<0$}
\STATE solve (P20) to update $(\widetilde{\boldsymbol{q}},\widetilde{\boldsymbol{\phi}})$;
\ENDIF
\STATE update $\boldsymbol{q}^{(t)}$ and $\boldsymbol{\phi}^{(t)}$ by (\ref{prb:stchs_phi_q_update});
\UNTIL{$\|\boldsymbol{\phi}^{(t)}\!-\!\boldsymbol{\phi}^{(t-1)}\|_2\!+\!\|\boldsymbol{q}^{(t)}\!-\!\boldsymbol{q}^{(t-1)}\|_2\leq\epsilon_0$}
\end{algorithmic}
\end{algorithm}
}

\section{Discussion}
\label{sec:disc}

\subsection{Extension}
\label{subsec:extn}

Although this paper studies weighted sum-power problem with SINR constraints, our proposed solutions can be easily extended to weighted sum/maximization of information rate, EE or min-SINR optimization. In fact, utilizing BCD methodology as well as appropriate transform, e.g. quadratic transform \cite{KMShen_2018} or Dinkelbach's method \cite{Dinkelbach_1967} and so on, all the above performance metrics can be equivalently transformed into quadratic problems. Via additionally possibly utilizing successive convex approximations (SCA), QoS constraints other than SINR, (e.g. energy transfer constraints, sum-rate, sum-power) can also be replaced by convex constraints (w.r.t. IRS phase-shifting). Consequently, the PDD/neADMM algorithms can again be invoked to design IRS, as illustrated previously.

\subsection{Complexity of PDD and neADMM}
\label{subsec:complexity}

Here we compare the complexity of PDD and neADMM methods. The major complexity of PDD comes from solving (P10) since all the other steps have analytic updating rules. Similarly, the primary complexity of neADMM lies in solving (P16). It is worth noting that the PDD method works well when the constraint (\ref{eq:feas_al_cnstr_phi_bd}) is omitted in practice (we present (\ref{eq:feas_al_cnstr_phi_bd}) in (P10) because it can simplify the convergence analysis in Theorem \ref{thm:pdd_cnvg}). According to Sec.6.2.2 of \cite{SDP_textbook_BenTal}, the complexity of solving (P16) or (P10), both of which are second order cone programming (SOCP) problems, is identical, i.e. $\mathcal{O}(K^{1.5}N^{3})$. As shown in Sec.\ref{sec:numerical_results}, the PDD method usually needs to solve the SOCP sub-problem more times than the neADMM due to its intrinsic two-layer structure.

\section{Numerical Results}
\label{sec:numerical_results}

\begin{figure}
\centering
\includegraphics[width=0.46\textwidth]{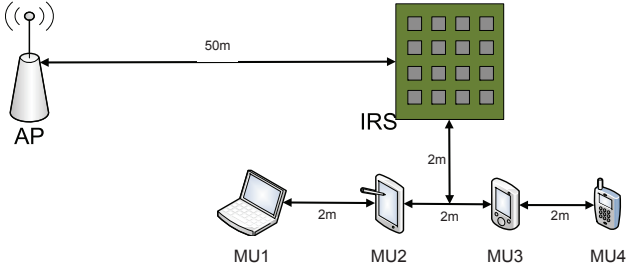} \vspace{-2pt}\caption{Simulation setup.}
\label{fig:sim_system}
\end{figure}

In this section, numerical results are presented to verify our proposed algorithms. As shown in Fig.~\ref{fig:sim_system}, a system that is similar to \cite{QQWu_TWC2019} comprises one AP with $M=4$ antennas, $4$ single-antenna MUs and one IRS. The number of reflecting elements $N$ varies from $30$ to $120$. The distance between the AP and IRS is $50$ meters (m) and it is assumed that a line-of-sight (LoS) propagation path between them exists. The $4$ MUs are located in a row parallel to the IRS-AP line and with a $2$m spacing between the adjacent peers. Signals emitting from MUs to both the AP and the IRS experience $10$dB penetration loss, independent Rayleigh fading and the pathloss exponent of $3$. The noise has -$180$dBm/Hz and the channel bandwidth is $100$KHz. Therefore $\sigma^2 = 10^{-13}$mW. We set the transmission power of each MU as $10$dBm and $5$dBm for the feasibility detection and the sum-power minimization problem, respectively. Besides, it is assumed that the antenna gain of both the AP and MUs is $0$dBi and that of each reflecting element at the IRS is $5$dBi.

\begin{figure}
\centering
\includegraphics[width=0.48\textwidth]{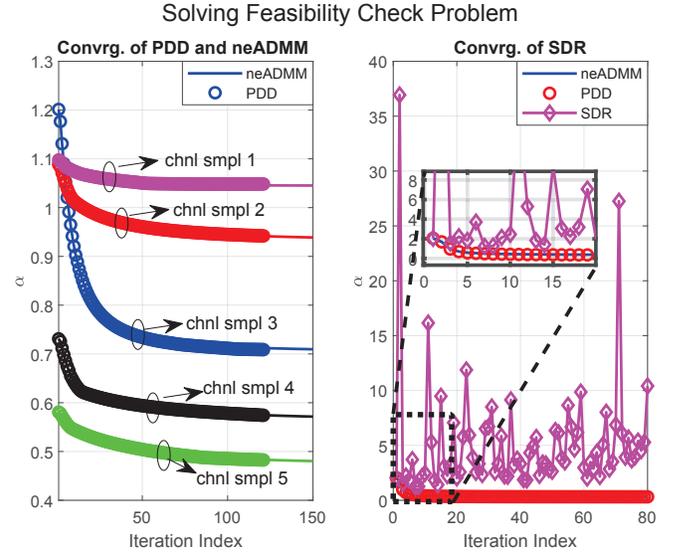}
\caption{Iterative detecting the feasibility.}
\label{fig:itinery_feasChk}
\end{figure}

Fig.~\ref{fig:itinery_feasChk} illustrates the behaviors of PDD, neADMM and SDR methods in checking feasibility. In our comparison, the PDD and neADMM algorithms always start from identical starting points. Interestingly, the PDD and neADMM solutions appear to exhibit concerted improvement in each iteration. Converge is usually reached within a hundred iterations. Besides, we also compare our proposed two algorithms with the classical SDR method in \cite{QQWu_TWC2019}. In the simulation, $10000$ Gaussian samples are generated to approximate rank-1 solution. As shown in Fig.~\ref{fig:itinery_feasChk}, the SDR method totally loses its power. The issue lies in that the element-wise normalization is performed to the Gaussian samples to satisfy the unit-modulus constraints. This operation destroys the optimal solution structure obtained by SDP, which severely deteriorates the objective value and renders the objective itineraries a random sequence. In our experiments, the SDR method generally does not converge for any channel realizations.

\begin{figure}
\centering
\includegraphics[width=0.48\textwidth]{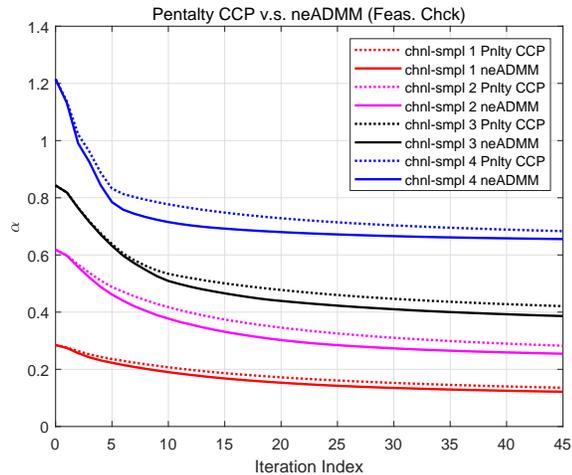}
\caption{Penalty CCP \cite{GuiZhou_Rbst_IRS_TSP,GuiZhou_Rbst_IRS_Lett} v.s. our solutions.}
\label{fig:ccp_vs_neadmm}
\end{figure}
\begin{figure}
\centering
\includegraphics[width=0.48\textwidth]{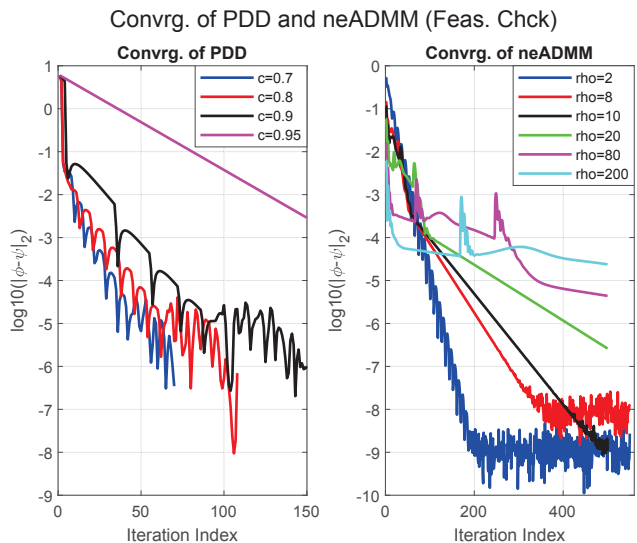} \vspace{-2pt}\caption{Convrg. of PDD and neADMM (feas. chck).}
\label{fig:cnvg_feasChck_neadmm_pdd}
\end{figure}

Fig.~\ref{fig:ccp_vs_neadmm} compares the penalty CCP method in \cite{GuiZhou_Rbst_IRS_TSP,GuiZhou_Rbst_IRS_Lett} and our solutions for feasibility check. Since PDD and neADMM exhibit identical convergence (as shown in Fig.\ref{fig:cnvg_feasChck_neadmm_pdd}), we only plot neADMM in Fig.~\ref{fig:ccp_vs_neadmm} for clearance. In the experiment, the initial penalty is set as $1$ and the penalty inflation coefficient ($\gamma$ of Alg.1 in \cite{GuiZhou_Rbst_IRS_Lett}) is set as $1.2$ (which generally yields the best converge according to experiments). As shown in Fig.~\ref{fig:ccp_vs_neadmm} ($N=60$), the penalty CCP and our solutions start from identical initial points for each specific channel realization. We find that our solutions always achieve better performance than the penalty CCP. Moreover, our solution is more efficient. In fact, compared with our (P10)\footnote{Note that the (\ref{eq:feas_al_cnstr_phi_bd}) in (P10) can be omitted in implementation} or (P16), the penalty CCP problem (14) \cite{GuiZhou_Rbst_IRS_Lett} has additional $2N$ linear constraints, which intensively enlarges the problem size ($N$ is usually large).

Next we examine the convergence characteristics of our proposed algorithms. Fig.~\ref{fig:cnvg_feasChck_neadmm_pdd} shows the impact of parameter $c$ on the convergence rate of PDD algorithm. Recall that $c$ is utilized to adjust the penalty. The lower value $c$ takes, the faster the algorithm converges. At the same time, however, when $c$ is larger, the penalty parameter inflates more smoothly and consequently the whole procedure presents more improvement in the objective value. In practice, the value from the range $[0.8,0.9]$ is usually a well-balanced choice for $c$. The impact of the penalty coefficient $\rho$ on neADMM is  also illustrated in Fig.\ref{fig:cnvg_feasChck_neadmm_pdd}. When $\rho$ is too large, the iterates will soon got stuck and stop improving. Generally a value in $[2,8]$ is an appropriate choice for $\rho$ to yield fast convergence.

\begin{figure}
\centering
\includegraphics[width=0.48\textwidth]{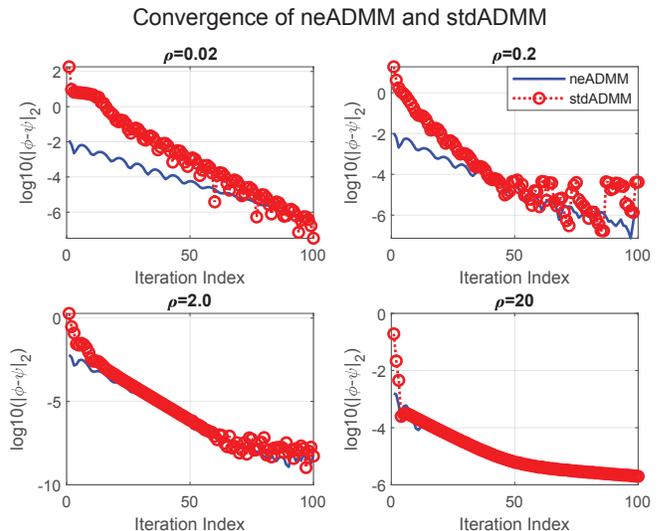} \vspace{-2pt}\caption{neADMM v.s. stdADMM.}
\label{fig:cnvg_neadmm_vs_stdadmm}
\end{figure}

We proceed to compare the standard ADMM (stdADMM) \cite{admm_Boyd_2011} and the novel neADMM algorithms. Fig.~\ref{fig:cnvg_neadmm_vs_stdadmm} presents the convergence behaviors of these two peers with the same penalty coefficient $\rho$ ranging from $0.02$ to $20$. The two counterparts solve the same problem starting from common initial points. As shown in Fig.~\ref{fig:cnvg_neadmm_vs_stdadmm}, stdADMM and neADMM generally have the same convergence rate in the long run in most test-cases. However one remarkable advantage of neADMM is that it usually exhibits higher precision in the first several tens of iterations, during which the algorithm can usually reach a precision of $10^{-\!3}$ to $10^{-\!4}$, which is indeed a rather satisfactory precision for practice. Therefore neADMM is more preferable to the stdADMM.

\begin{table}[htbp]
\centering
\label{tab:complexity}
\caption{Complexity of PDD and neADMM}
\begin{tabular}{|c|c|c|c|c|}
\hline
\multirow{2}{*}{} & \multicolumn{2}{|c|}{PDD} & \multicolumn{2}{|c|}{neADMM} \\
\cline{2-5}
\multirow{2}{*}{} & Time (Sec.) & \# (P10) & Time (Sec.) & \# (P16) \\
\hline
$N=60$ & 16.21 & 44.38 & 12.55 & 32.25 \\
\hline
$N=120$ & 18.16 & 47.33 & 17.67 & 42.00 \\
\hline
$N=200$ & 19.25 & 47.58 & 16.63 & 38.78 \\ 
\hline
\end{tabular}
\end{table}

In Table I we compare the complexity of PDD and neADMM, whose major complexity lies in repeatedly solving the subproblems (P10) and (P16), respectively. In the experiment, we invoke both algorithms to solve the same problem from identical starting points and use the same stopping criterion, i.e. $\|\boldsymbol{\phi}^{(t)}\!-\!\boldsymbol{\phi}^{(t\!-\!1)}\|_2^2<10^{-4}$ and $\|\boldsymbol{q}^{(t)}\!-\!\boldsymbol{q}^{(t\!-\!1)}\|_2^2<10^{-4}$, to terminate. For the PDD method, the inner layer stopping criterion is that the objective improvement relative to the objective value is smaller than $10^{-4}$. Table~I shows the average running time and the number of times of (P10) or (P16) being solved to accomplish $\boldsymbol{\phi}$ update. Empirically the neADMM is more efficient than the PDD method.

\begin{figure}
\centering
  \includegraphics[width=0.48\textwidth]{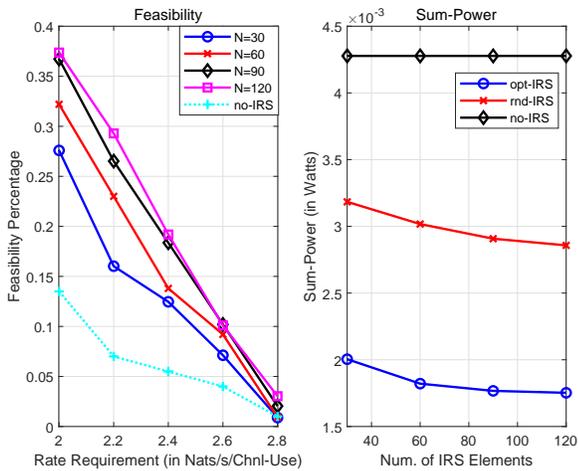} \vspace{-2pt}\caption{IRS improves the feasibility and sum-power}.
 \label{fig:feas_pwr_imprv}
\end{figure}

Fig.~\ref{fig:feas_pwr_imprv} illustrates the impact of IRS on feasibility and sum-power. In the experiment, the QoS requirements of all MUs are identical. For each specific value of information rate requirement and $N$, $200$ random channel realizations are generated and Alg.\ref{alg:feas_check} is invoked to find a feasible solution. The percentage of the channel realizations whose feasible solution can be found are plotted. We also present the benefits brought by IRS in power saving ($\boldsymbol{\kappa}=\mathbf{1}_K$). $400$ channel realizations are generated and Alg.~\ref{alg:opt-overall} is employed. The resultant average sum-power is presented. Clearly Fig.~\ref{fig:feas_pwr_imprv} demonstrates that IRS significantly improves the feasibility and sum-power performance.

\begin{figure}
\centering
\includegraphics[width=0.48\textwidth]{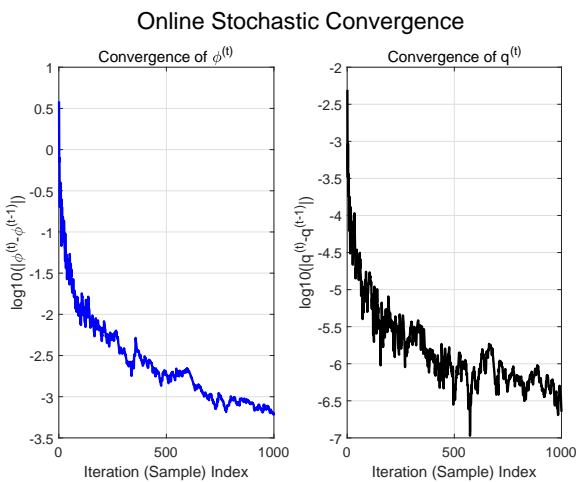} \vspace{-2pt}
\caption{Stochastic online algorithm converges.}
\label{fig:stchs_cnvg}
\end{figure}

\begin{figure}
\centering
\includegraphics[width=0.48\textwidth]{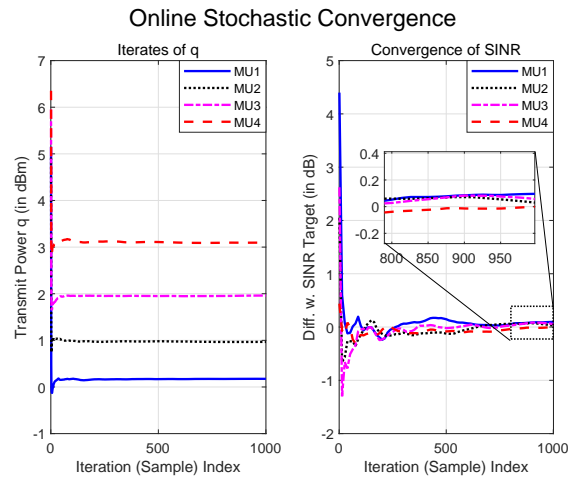} \vspace{-2pt}
\caption{Optimization results by the stochastic online algorithms.}
\label{fig:stchs_prfrmn}
\end{figure}
We continue to examine the performance of the stochastic online algorithm when CSI is imperfect. In our simulation, the IRS-AP link $\boldsymbol{G}$ is assumed to be LoS and perfectly known \cite{IRS_ChnEst_LDai} \footnote{In fact, even if the IRS-AP link is random, Alg.~\ref{alg:stchs-pwr_min} still converges at an expense of more samples.}. We assume that all the CSI estimates $\widehat{\boldsymbol{h}}_{i,k}$, $i\in\{\mathrm{r},\mathrm{d}\}$ $\forall k$, follow Rayleigh distribution with variance $\sigma^2_{i,k}\mathbf{I}_M$. Note this setting is equivalent to the generic model $\widehat{\boldsymbol{h}}_{i,k}\!=\!\boldsymbol{h}_{i,k}\!+\!\boldsymbol{\Delta}_{i,k}$ with $\boldsymbol{h}_{i,k}$ and $\boldsymbol{\Delta}_{i,k}$ both following Gaussian distribution with variance $\epsilon^2_{i,k}\mathbf{I}_M$ and $\tau^2_{i,k}\mathbf{I}_M$, respectively, such that $\epsilon^2_{i,k}\!=\!\eta\sigma^2_{i,k}$ and $\tau^2_{i,k}\!=\!(1\!-\!\eta)\sigma^2_{i,k}$ $\forall\eta\in[0,1]$. The learning parameter is chosen as $\rho^{(t)}\!=\!3/(t\!+\!1)^{0.9}$. Fig.~\ref{fig:stchs_cnvg} shows the convergence property of Alg.~\ref{alg:stchs-pwr_min}. Fig.~\ref{fig:stchs_prfrmn} illustrates the optimization effects. In the experiment, we impose an uniform SINR constraint of $11.5$dB to all MUs. As shown in Fig.~\ref{fig:stchs_cnvg}, the result transmit power increases with the AP-MU distance, which coincides with intuition. Fig.~\ref{fig:stchs_cnvg} also shows the difference between the iterates' stochastic SINR and its SINR target ($11.5$dB). As can been seen, the optimized IRS configuration and power yield tight SINR constraints in a stochastic sense, which coincides with the classical power allocation result for perfect CSI.

\section{Conclusion}
\label{sec:conclusion}

In this paper we perform a comprehensive research on the power minimization problem in an IRS-aided multi-user MISO uplink wireless network with SINR constraints. We first study the feasibility problem and provide a new sufficient condition ensuring arbitrary SINR requirements. Besides, we propose novel methods to resolve the IRS-dependent-QoS-constraints for IRS design. These new methods have successfully solve a family of IRS optimization problems in both perfect and imperfect CSI settings. As demonstrated, our proposed algorithms outperform existing methods in both performance and complexity and can easily extend to a wider range of IRS optimization tasks. 

As a possible future extension of this work, we can also consider IRS design under various IRS-dependent-QoS-constraints beyond the SINR constraints. For instance, we can impose requirement(s) of sum-rate, sum-power, energy harvesting (i.e. the power harvested at the energy receiver should be larger than a threshold) and energy sensing (i.e. the primary user's sensed energy from different secondary users should be lower than some threshold in CR) on our optimization problems. Via possibility invoking quadratic transform \cite{KMShen_2018} or SCA method, these aforementioned constraints can be transformed into convex constraints with respect to the IRS phase-shifters. Then our proposed PDD/neADMM solutions can again be invoked straightforwardly.

\appendix

\subsection{Proof of Lemma \ref{lem:feas_cond}}
\label{subsec:prof_feas_cond}
\begin{proof}
Assume that the nonzero singular values of an arbitrary matrix $\boldsymbol{A}$ is $\sigma_1(\boldsymbol{A})\geq\cdots\geq\sigma_r(\boldsymbol{A})$, with $r$ being the rank of $\boldsymbol{A}$,  
Then according to relation between singular value decomposition and low rank approximation \cite{SVD_lowrank}, for any $k<r$, we have 
\begin{shrinkeq}{-1ex}
\begin{align}
\sigma_{k+1}(\boldsymbol{A}) = \min_{\mathsf{rank}(\boldsymbol{B})=k}\|\boldsymbol{A}-\boldsymbol{B}\|_2, \label{eq:singular_lowrank}
\end{align}
\end{shrinkeq}

Due to the rank assumption $\mathsf{rank}(\boldsymbol{H}_{\mathrm{d}})=K$. Define $\boldsymbol{\Delta}\triangleq\boldsymbol{G}\boldsymbol{\Phi}\boldsymbol{H}_{\mathrm{r}}$. We prove that $\mathsf{rank}(\boldsymbol{H})=K$ by contradiction. If $\mathsf{rank}(\boldsymbol{H})\!=\!\mathsf{rank}(\boldsymbol{H}_{\mathrm{d}}\!+\!\boldsymbol{\Delta})\!=\!r\!<\!K$, then according to the relation in (\ref{eq:singular_lowrank}), we have
\begin{shrinkeq}{-1ex}
\begin{align}
\|\boldsymbol{\Delta}\|_2 &= \|\boldsymbol{H}_{\mathrm{d}}-\big(\boldsymbol{H}_{\mathrm{d}}+\boldsymbol{\Delta}\big)\|_2 \label{eq:rank_left}\\
&\geq\min_{\mathsf{rank}(\boldsymbol{B})=r}\|\boldsymbol{H}_{\mathrm{d}}-\boldsymbol{B}\|_2 = \sigma_{r\!+\!1}\big(\boldsymbol{H}_{\mathrm{d}}\big)\geq\sigma_{\min}\big(\boldsymbol{H}_{\mathrm{d}}\big). \nonumber 
\end{align}
\end{shrinkeq}
At the same time, however, by property of matrix norm we have
\begin{shrinkeq}{-1ex}
\begin{align}
\!\!\!\!\|\boldsymbol{\Delta}\|_2 = \|\boldsymbol{G}\boldsymbol{\Phi}\boldsymbol{H}_{\mathrm{r}}\|_2 & \leq \|\boldsymbol{G}\|_2\|\boldsymbol{\Phi}\|_2\|\boldsymbol{H}_{\mathrm{r}}\|_2 \label{eq:rank_right}\\
\!\!\!\!&\overset{(a)}{=}\sigma_{\max}(\boldsymbol{G})\cdot\sigma_{\max}(\boldsymbol{H}_{\mathrm{r}})<\sigma_{K}(\boldsymbol{H}_{\mathrm{d}}), \nonumber 
\end{align}
\end{shrinkeq}
where the step $(a)$ utilizes the fact that $\|\boldsymbol{\Phi}\|_2=1$ and the last inequality follows the assumption of the lemma. Combining (\ref{eq:rank_left}) and (\ref{eq:rank_right}), a contradiction has been reached. Therefore we conclude $\mathsf{rank}(\boldsymbol{H})=K$, i.e. the effective channel $\boldsymbol{H}$ is full rank. Invoking Theorem 3.1 of \cite{complete_QoS_RHunger}, the assertion in the lemma is proved. 
\end{proof}

\subsection{Proof of Theorem \ref{thm:pdd_cnvg}}
\label{subsec:prof_pdd_cnvg}
\begin{proof}
For simplicity, define $g_k(\boldsymbol{\psi},\alpha)\triangleq\boldsymbol{\phi}^H\boldsymbol{Q}_k\boldsymbol{\phi}\!+\!2\mathsf{Re}\{\boldsymbol{q}_k^H\boldsymbol{\phi}\}\!+\!d_k\!-\!\alpha$, $\forall k\in\mathcal{K}$ and rewrite the constraints (\ref{eq:feas_prob_rate_cnstr}) as $g_k(\boldsymbol{\psi},\alpha)\leq0$, $\forall k\in\mathcal{K}$. Besides, denote the manifold $\mathcal{M}\triangleq\{\boldsymbol{\psi}\ \text{s.t.}\ |\boldsymbol{\psi}|=\mathbf{1}_N\}$. 

Since $\{\boldsymbol{\phi}^{(k)}\}$ is bounded by the constraint (\ref{eq:feas_prob_bd_cnstr}), and $\boldsymbol{\psi}$ and $\alpha$ are actually updated via continuous function of $\boldsymbol{\phi}$ (in fact $\alpha$ is obtained by $\alpha=\max_{k\in\mathcal{K}}\{\boldsymbol{\phi}^H\boldsymbol{Q}_k\boldsymbol{\phi}\!+\!2\mathsf{Re}\{\boldsymbol{q}_k^H\boldsymbol{\phi}\}\!+\!d_k\}$, which is a continuous function in $\boldsymbol{\phi}$). Therefore there exists limit point(s) of the solution iterates. Without loss of generality, we just suppose that $\big(\boldsymbol{\phi}^{(k)},\boldsymbol{\psi}^{(k)},\alpha^{(k)}\big)$ is a limit point, i.e. $\big(\boldsymbol{\phi}^{(k)},\boldsymbol{\psi}^{(k)},\alpha^{(k)}\big)\rightarrow\big(\widebar{\boldsymbol{\phi}},\widebar{\boldsymbol{\psi}},\widebar{\alpha}\big)$.

For any $k\in\{1,2,\cdots\}$, assume that $\big(\boldsymbol{\phi}^{(k)},\boldsymbol{\psi}^{(k)},\alpha^{(k)}\big)$ is limit of the convergent subsequence $\big(\boldsymbol{\phi}^{(k,t_j)},\boldsymbol{\psi}^{(k,t_j)},\alpha^{(k,t_j)}\big)$ of $\big(\boldsymbol{\phi}^{(k,t)},\boldsymbol{\psi}^{(k,t)},\alpha^{(k,t)}\big)$. Denote the objective function (P9) is $h(\boldsymbol{\phi},\boldsymbol{\psi},\alpha)$. Since the inner layer of Alg.\ref{alg:PDD_alg} is 2-block BCD procedure, the objective sequence $h^{(k,t)}$ is decreasing, bounded from below and consequently convergent. Suppose that (by possibly restricting to a subsequence) $\boldsymbol{\psi}^{(k,t_j\!-\!1)}\rightarrow\widehat{\boldsymbol{\psi}}$. Since $\boldsymbol{\psi}^{(k,t_j)}=\mathsf{argmin}_{\boldsymbol{\psi}\in\mathcal{M}}h(\boldsymbol{\phi}^{(k,t_j)},\boldsymbol{\psi},\alpha^{(k,t_j)})$, so
\begin{align}
h(\boldsymbol{\phi}^{(k,t_j)},\boldsymbol{\psi}^{(k,t_j)},\alpha^{(k,t_j)})\leq h(\boldsymbol{\phi}^{(k,t_j)},\boldsymbol{\psi},\alpha^{(k,t_j)}),\ \forall \boldsymbol{\psi}\in\mathcal{M}. \nonumber
\end{align}
Taking $j\rightarrow\infty$ in the above inequality, we obtain 
\begin{align}
h(\boldsymbol{\phi}^{(k)},\boldsymbol{\psi}^{(k)},\alpha^{(k)})\leq  h(\boldsymbol{\phi}^{(k)},\boldsymbol{\psi},\alpha^{(k)}),\ \forall \boldsymbol{\psi}\in\mathcal{M}. \nonumber
\end{align}
Hence $\boldsymbol{\psi}^{(k)}$ is the optimal solution to the problem $\min_{\boldsymbol{\psi}\in\mathcal{M}}h(\boldsymbol{\phi}^{(k)},\boldsymbol{\psi},\alpha^{(k)})$. At the same time, note that the objective sequence $\{h^{(k,t)}\}_{t\!=\!1}^{\infty}$ converges, we have 
\begin{align}
h(\boldsymbol{\phi}^{(k)},\boldsymbol{\psi}^{(k)},\alpha^{(k)}) = h(\boldsymbol{\phi}^{(k)},\widehat{\boldsymbol{\psi}},\alpha^{(k)}). 
\end{align}
Therefore ${\widehat{\boldsymbol{\psi}}}$ is also an optimal solution to the problem $\min_{\boldsymbol{\psi}\in\mathcal{M}}h(\boldsymbol{\phi}^{(k)},\boldsymbol{\psi},\alpha^{(k)})$. However noticing the fact that this optimization problem has unique solution (recall the argument obtaining (\ref{eq:psi_update})), we conclude $\boldsymbol{\psi}^{(k)}=\widehat{\boldsymbol{\psi}}$. 

Note that $(\boldsymbol{\phi}^{(k,t_j)},\alpha^{(k,t_j)})$ is the solution to the problem $\min_{g_i(\boldsymbol{\phi},\alpha)\leq0,|\boldsymbol{\phi}|\leq\mathbf{1}_N}h(\boldsymbol{\phi},\boldsymbol{\psi}^{(k,t_j\!-\!1)},\alpha)$. Also note that the problem $\min_{g_i(\boldsymbol{\phi},\alpha)\leq0,|\boldsymbol{\phi}|\leq\mathbf{1}_N}h(\boldsymbol{\phi},\boldsymbol{\psi}^{(k,t_j\!-\!1)},\alpha)$ satisfies Slater's condition (since $\boldsymbol{\phi}$ can be chosen suficiently small and $\alpha$ can be chosen sufficiently large) and therefore it satisfies Robinson's condition (Sec.V.A in \cite{QJShi_PDD_org}). According to Lemma 3.26 of \cite{NP_ARuszczynski}, the Lagrangian multipliers $\boldsymbol{\nu}^{(k,t_j)}$ and $\boldsymbol{\varpi}^{(k,t_j)}$ are bounded. By checking its KKT condition of the problem $\min_{g_i(\boldsymbol{\phi},\alpha)\leq0,|\boldsymbol{\phi}|\leq\mathbf{1}_N}h(\boldsymbol{\phi},\boldsymbol{\psi}^{(k,t_j\!-\!1)},\alpha)$, taking the limit $j\rightarrow\infty$, substituting $\widehat{\boldsymbol{\psi}}$ with $\boldsymbol{\psi}^{(k)}$ in the limit and restricting to a subsequence of $\{\boldsymbol{\nu}^{(k,t_j)}\}$ and $\{\boldsymbol{\varpi}^{(k,t_j)}\}$, we obtain the KKT conditions at $(\boldsymbol{\phi}^{(k)},\boldsymbol{\psi}^{(k)},\alpha^{(k)})$ as follows:
\begin{subequations}
\begin{align}
&\frac{1}{2\rho^{(k)}}\big(\boldsymbol{\phi}^{(k)}\!-\!\boldsymbol{\psi}^{(k)}\big)\!+\!\frac{1}{2}\boldsymbol{\lambda}^{(k)}\!+\!\sum_{i\in\mathcal{K}}{\nu}_i^{(k)}\frac{\partial g_i}{\partial \boldsymbol{\phi}^{(k)*}} \\ 
&\qquad\qquad\qquad\qquad\qquad\  + \sum_{n\in\mathcal{N}}\varpi_n^{(k)}\frac{\partial |\boldsymbol{\phi}_n^{(k)}|}{\partial \boldsymbol{\phi}^{(k)*}}=\mathbf{0}, \nonumber \\
&\ \  1-\sum_{i\in\mathcal{K}}\nu_i^{(k)} = 0, \\
&g_i(\boldsymbol{\phi}^{(k)}\!,\!\alpha^{(k)})\leq0,\nu_i^{(k)}g_i(\boldsymbol{\phi}^{(k)}\!,\!\alpha^{(k)})\!=\!0,\nu_i^{(k)}\!\geq\!0,\forall i,\\
&|\phi_n^{(k)}|\leq1,\varpi_n^{(k)}(|\phi_n^{(k)}|\!-\!1)=0,\ \varpi_n^{(k)}\geq0,\ \forall n\in\mathcal{N}. 
\end{align}
\end{subequations}

Denote $\boldsymbol{\mu}^{(k)}\triangleq\frac{1}{\rho^{(k)}}\big(\boldsymbol{\phi}^{(k)}-\boldsymbol{\psi}^{(k)}\big)+\boldsymbol{\lambda}^{(k)}$. By assumption $\boldsymbol{\mu}^{(k)}$ is bounded. Therefore by taking $k\rightarrow\infty$ and restricting to a subsequence of $\{\boldsymbol{\nu}^{(k)}\}$ and $\{\boldsymbol{\varpi}^{(k)}\}$, we obtain 
\begin{subequations}
\label{eq:kkt_x}
\begin{align}
&\frac{\partial}{\partial \bar{\boldsymbol{\phi}}^{*}}\mathsf{Re}\big\{\bar{\boldsymbol{\mu}}^H \big(\bar{\boldsymbol{\phi}}\!-\!\bar{\boldsymbol{\psi}}\big)\big\}+\sum_{i\in\mathcal{K}}\bar{\nu_i}\frac{\partial g_i}{\partial \bar{\boldsymbol{\phi}}^{*}} \nonumber\\
&\qquad\qquad\qquad + \sum_{n\in\mathcal{N}}\bar{\varpi}_n\frac{\partial}{\partial\bar{\boldsymbol{\phi}}^{*}}\big(|\bar{\boldsymbol{\phi}}_n^{(k)}|-1\big)=\mathbf{0}, \\
& 1-\sum_{i\in\mathcal{K}}\bar{\nu}_i = 0, \\
&g_i(\bar{\boldsymbol{\phi}},\bar{\alpha})\leq0,\  \bar{\nu}_ig_i(\bar{\boldsymbol{\phi}},\bar{\alpha})=0,\ \bar{\nu}_i\geq0,\ \forall i\in\mathcal{K},\\
&|\bar{\phi}_n|\leq1, \ \bar{\varpi}_n(|\bar{\phi}_n|\!-\!1)=0, \ \bar{\varpi}_n\geq0,\ \forall n\in\mathcal{N}. 
\end{align}
\end{subequations}

Besides, by assumption, $\boldsymbol{\mu}^{(k)}$ is bounded and therefore $\boldsymbol{\lambda}^{(k)}$ is bounded. According to steps 9-13 of Alg.\ref{alg:PDD_alg}, at least one of the two possible cases is performed infinitely many times. That is---either i): $\rho^{(k)}\rightarrow0$ with $(\boldsymbol{\mu}^{(k)}\!-\!\boldsymbol{\lambda}^{(k)})$ bounded, or ii): $(\boldsymbol{\mu}^{(k)}\!-\!\boldsymbol{\lambda}^{(k)})\rightarrow0$ with $\rho^{(k)}$ bounded. Hence we always have  
\begin{align}
\boldsymbol{\phi}^{(k)}-\boldsymbol{\psi}^{(k)}=\rho^{(k)}\big(\boldsymbol{\mu}^{(k)}-\boldsymbol{\lambda}^{(k)}\big)\rightarrow0. \label{eq:equality_stands}
\end{align}
Notice that $\{\mu^{(k)}\}$ is bounded. Then taking $k$ to infinity and restricting to a subsequence of $\{\mu^{(k)}\}$, we obtain
\begin{align}
\!\!\bar{\boldsymbol{\phi}}\!-\!\bar{\boldsymbol{\psi}}\!=\!\mathbf{0},\mathsf{Re}\big\{\bar{\boldsymbol{\mu}}^*\!\circ\!\big(\bar{\boldsymbol{\phi}}\!-\!\bar{\boldsymbol{\psi}}\big)\big\}\!=\!\mathbf{0},\mathsf{Re}\big\{\bar{\boldsymbol{\mu}}\!\circ\!\big(\bar{\boldsymbol{\phi}}\!-\!\bar{\boldsymbol{\psi}}\big)\big\}\!=\!\mathbf{0},\label{eq:kkt_x_3}
\end{align}
where the notation ``$\circ$'' means element-wise production. 

At the same time, when $\big(\boldsymbol{\phi}^{(k,t_j)},\alpha^{(k,t_j)}\big)$ are given, $\boldsymbol{\psi}$ is updated by solving $\textup{P(11)}^{\prime}$. According to previous discussion, the $\boldsymbol{\psi}^{(k,t_j)}$ is an optimal solution to $\textup{P(11)}^{\prime}$ if and only if it is optimal to $\textup{P(11)}$, whose KKT condition reads
\begin{subequations}
\begin{align}
&-\frac{1}{2\rho^{(k)}}\big(\boldsymbol{\phi}^{(k,t_j)}\!-\!\boldsymbol{\psi}^{(k,t_j)}\big)-\frac{1}{2}\boldsymbol{\lambda}^{(k)}\label{eq:kkt_y_1}\\
& \qquad\qquad\qquad + \sum_{n\in\mathcal{N}}\xi_{n}^{(k,t_j)}\frac{\partial}{\partial \boldsymbol{\psi}^{(k,t_j)*}}\Big(|\psi_{n}^{(k)}|\!-\!1\Big)=\mathbf{0}, \nonumber \\
&\quad \boldsymbol{\xi}^{(k,t_j)}\circ\big(|\boldsymbol{\psi}^{(k,t_j)}\!-\!\mathbf{1}_N|\big)=\mathbf{0},\ \ \boldsymbol{\psi}^{(k,t_j)}=\mathbf{1}_N. 
\end{align}
\end{subequations}
In fact, by (\ref{eq:kkt_y_1}) we can obtain 
\begin{align}
\!\!\!\!\!\boldsymbol{\xi}^{(k,t_j)}\!=\!\big(\boldsymbol{\psi}^{(k,t_j)}\big){}^{-\!1}\!\circ\!\big[\big(\rho^{(k)}\big){}^{-\!1}\big(\boldsymbol{\phi}^{(k,t_j)}\!\!-\!\!\boldsymbol{\psi}^{(k,t_j)}\big)\!+\!\boldsymbol{\lambda}^{(k)}\big], 
\end{align}
which is a continuous function in $\boldsymbol{\phi}^{(k,t_j)}$ and $\boldsymbol{\psi}^{(k,t_j)}$. By firstly taking $j\rightarrow\infty$, then $k\rightarrow\infty$ and restricting to a subsequence of $\{\boldsymbol{\xi}^{(k)}\}$, we obtain 
\begin{subequations}
\label{eq:kkt_y_final}
\begin{align}
&\frac{\partial}{\partial \bar{\boldsymbol{\psi}}^{*}}\mathsf{Re}\big\{\bar{\boldsymbol{\mu}}^H \big(\bar{\boldsymbol{\phi}}\!-\!\bar{\boldsymbol{\psi}}\big)\big\} 
+ \sum_{n\in\mathcal{N}}\bar{\xi}_{n}\frac{\partial}{\partial \bar{\boldsymbol{\psi}}^{*}}\Big(|\bar{\psi}_{n}|\!-\!1\Big)=\mathbf{0}. \label{eq:kkt_y_final_1}\\
&\qquad \bar{\boldsymbol{\xi}}\circ\big(|\bar{\boldsymbol{\psi}}-\mathbf{1}_N|\big)=\mathbf{0}, \ \ \bar{\boldsymbol{\psi}}=\mathbf{1}_N. \label{eq:kkt_y_final_2}
\end{align}
\end{subequations}
Combining the equations (\ref{eq:kkt_x}), (\ref{eq:kkt_x_3}) and (\ref{eq:kkt_y_final}), we actually find the Lagrangian multipliers $(\widebar{\boldsymbol{\nu}},\widebar{\boldsymbol{\varpi}},\widebar{\boldsymbol{\mu}},\widebar{\boldsymbol{\xi}})$ satisfying the exact KKT conditions of (P8), with the complex vector $\bar{\boldsymbol{\mu}}$ being the Lagrangian multipliers associated with the equality constraint (\ref{eq:feas_prob_xy}).  The proof is complete.
\end{proof}

\subsection{Proof of Lemma \ref{lem:psi_solution}}
\label{subsec:prof_psi_solution}
\begin{proof}
First, it can be readily recognized that the problem (P17) can be decomposed into $N$ independent separate subproblems with each problem dealing with only the $i$-th entry  $\psi_i$ of $\boldsymbol{\psi}$, $\forall i\in\mathcal{N}$, which is given as 
\begin{align}
(\textup{P}17_i): \min_{\psi_i} \big|\psi_i\!-\!\phi_i\!+\!\rho^{\!-\!1}z_{2,i}\big|^2 + \big| |\psi_i|\!-\!1\!+\!\rho^{\!-\!1}z_{1,i}\big|^2
\end{align}

After some manipulation, the above problem can be rewritten as follows
\begin{align}
\!\!\!\!\!\!(\textup{P}18_i): \min_{\psi_i} 2|\psi_i|^2 - 2\mathsf{Re}\{\psi_i^*s_i\}-2t_i|\psi_i|\triangleq g(\psi_i,\psi_i^*). 
\end{align}
with the coefficients $s_i$ and $t_i$ defined as
\begin{align}
s_i\triangleq\phi_i-\rho^{\!-\!1}z_{2,i}, \ t_i\triangleq\mathsf{Re}\{1-\rho^{-1}z_{1,i}\} 
\end{align}

According to Weierstrass' Theorem (Prop. 2.1.1 \cite{Book_BNO2003}), since the objective of (P18$_i$) is proper, continuous (and consequently closed ) and coercive, its minima exists. Therefore, by optimality condition, the minima $\psi_i^{\star}$ of (P18$_i$) should satisfy $\frac{\partial g}{\partial(\psi_i^{\star})^{*}}=0$. 

Besides, utilizing the chain rule for differential and noticing the face that 
\begin{shrinkeq}{-1ex}
\begin{subequations}
\begin{align}
\mathrm{d}|\psi_i| &= \mathrm{d}\big(\sqrt{|\psi_i|^2}\big) = \frac{1}{2|\psi_i|}\mathrm{d}\big(|\psi_i|^2\big) \\ 
& = \frac{1}{2|\psi_i|}\big(\psi_i\mathrm{d}{\psi_i^*}+\psi_i^*\mathrm{d}\psi_i\big), 
\end{align}
\end{subequations}
\end{shrinkeq}
we readily obtain that 
\begin{align}
\frac{\partial \big|\psi_i\big|}{\partial \psi_i^*} = \frac{1}{2\big|\psi_i\big|}\psi_i, \quad \frac{\partial \big|\psi_i\big|}{\partial \psi_i} = \frac{1}{2\big|\psi_i\big|}\psi_i^* .
\label{eq:derivative_abs}
\end{align}

Utilizing the (\ref{eq:derivative_abs}), the condition $\frac{\partial g}{\partial(\psi_i^{\star})^{*}}=0$ reads as follows
\begin{shrinkeq}{-1ex}
\begin{align}
2\psi_i^{\star}-s_i-\frac{t_i}{\big|\psi_i^{\star}\big|}\psi_i^{\star} = 0, 
\end{align}
\end{shrinkeq}
which implies 
\begin{shrinkeq}{-1ex}
\begin{align}
\psi_i^{\star}=\frac{s_i}{2-t_i/\big|\psi_i^{\star}\big|}\label{eq:psi_sol}
\end{align}
\end{shrinkeq}

Notice that $\psi_i^{\star}$ can be determined by (\ref{eq:psi_sol}) only if the $\big|\psi_i^{\star}\big|$ in the right hand side is known. To evaluate $\big|\psi_i^{\star}\big|$, we take modulus of both sides of (\ref{eq:psi_sol}) and obtain: 
\begin{align}
\big|2|\psi_i^{\star}|-t_i\big|=\big|s_i\big|. 
\end{align}
Therefore the $\big|\psi_i^{\star}\big|$ can be obtained as follows:
\begin{align}
\big|\psi_i^{\star}\big| = \frac{1}{2}\big(t_i\pm|s_i|\big). \label{eq:psi_modulus}
\end{align}

It is worth noting that we can obtain two possible values of $\big|\psi_i^{\star}\big|$ by (\ref{eq:psi_modulus}). According the previous discussion, since the true optimal solution $\psi_i^{\star}$ is always a solution to (\ref{eq:psi_sol}), therefore, at least one value of $\big|\psi_i^{\star}\big|$ in (\ref{eq:psi_modulus}) is non-negative. Then the determination of $\big|\psi_i^{\star}\big|$ and $\psi_i^{\star}$ is analyzed in the subsequent three cases:

\begin{itemize}
\item[C1)] if only one value (i.e. $t_i+|s_i|$) of $\big|\psi_i^{\star}\big|$ given by (\ref{eq:psi_modulus}) is positive, then it is indeed the modulus of real optimal $\psi_i^{\star}$. Substituting it back to (\ref{eq:psi_modulus}) we can determine the optimal $\psi_i^{\star}$. 

\item[C2)] if both values of $\big|\psi_i^{\star}\big|$ given by (\ref{eq:psi_modulus}) are positive, by substituting them back to (\ref{eq:psi_modulus}) we can obtain two candidates of $\psi_i^{\star}$ . Then we verify whether the modulus of the two candidates coincide with the their associated modulus value given in (\ref{eq:psi_modulus}). If one candidate fails the verification, the other candidate is the true value of $\psi_i^{\star}$.

\item[C3)] if both candidates obtained in C2) are feasible, then we choose the $\psi_i^{\star}$ giving the smaller objective value of (P18$_i$).  
\end{itemize}
The assertion has been proved. 
\end{proof}

\printbibliography

\end{document}